\documentclass[preprint*]{JHEP3} 

\JHEPspecialurl{http://jhep.sissa.it/JOURNAL/JHEP3.tar.gz}

\usepackage{epsfig,multicol}
\usepackage{amsmath,amssymb,bm}

\newcommand\fverb{\setbox\pippobox=\hbox\bgroup\verb}
\newcommand\fverbdo{\egroup\medskip\noindent%
                        \fbox{\unhbox\pippobox}\ }
\newcommand\fverbit{\egroup\item[\fbox{\unhbox\pippobox}]}
\newbox\pippobox

\title{
A simple construction of fermion measure term \\
in U(1) chiral lattice gauge theories \\
with exact gauge invariance
}

\author{Daisuke Kadoh \\ Center for Computational Sciences, University of Tsukuba, 
Ibaraki  305-8571, Japan\\
E-mail: \email{kadoh@ccs.tsukuba.ac.jp}}
\author{Yoshio Kikukawa \\ Institute of Physics, University of Tokyo, Tokyo 153-8902, Japan \\
E-mail: \email{kikukawa@hep1.c.u-tokyo.ac.jp}}

\preprint{
                 UT-KOMABA/07-16\\September 2007}      

\abstract{
In the gauge invariant formulation of  U(1) chiral lattice gauge theories based on 
the Ginsparg-Wilson relation,  
the gauge field dependence of the fermion measure is determined through the so-called measure term. 
We derive a closed formula of the measure term on the finite volume lattice. 
The Wilson line degrees of freedom (torons) of the link field are treated separately to take care of the global integrability. The local counter term is explicitly constructed with the local current associated with the cohomologically trivial part of  the gauge anomaly in finite volume. 
The resulted formula is very close to the known expression of the measure term in infinite volume 
with a single parameter integration,  and would be useful in practical implementations. 
}


\begin{document} 


\section{Introduction}
\label{sec:intro} 
Chiral gauge theories have several interesting possibilities in their own dynamics: 
fermion number non-conservation due to chiral anomaly\cite{'tHooft:1976up, 'tHooft:1976fv}, 
various realizations of the gauge symmetry and global flavor symmetry\cite{Raby:1979my, Dimopoulos:1980hn},  
the existence of massless composite fermions suggested by 't Hooft's anomaly matching condition\cite{'tHooft:1979bh} and so on. 
Unfortunately,  very little is known so far about the actual behavior of chiral gauge theories beyond
perturbation theory.  It is desirable to develop a formulation to study  
the non-perturbative dynamics of chiral gauge theories. 

Lattice gauge theory can now provide
a framework for non-perturbative formulation of  chiral gauge theories. 
The clue to this development is 
the construction of gauge-covariant and local lattice Dirac operators 
satisfying the Ginsparg-Wilson relation\cite{Ginsparg:1981bj,
Neuberger:1997fp,Hasenfratz:1998ri,Neuberger:1998wv,
Hasenfratz:1998jp,Hernandez:1998et}.\footnote{
An explicit solution of the Ginsparg-Wilson relation was derived from 
the overlap formalism proposed by Narayanan and Neuberger\cite{
Narayanan:wx,Narayanan:sk,Narayanan:ss,Narayanan:1994gw,Narayanan:1993gq,
Neuberger:1999ry,Narayanan:1996cu,Huet:1996pw,
Narayanan:1997by,Kikukawa:1997qh,Neuberger:1998xn} and is referred as the overlap Dirac operator. The overlap formalism gives a well-defined partition function of Weyl fermions on the lattice,
which nicely reproduces the fermion zero mode and the fermion-number
violating observables
('t Hooft vertices)\cite{Narayanan:1996kz,Kikukawa:1997md,Kikukawa:1997dv}. 
Through the recent re-discovery of the Ginsparg-Wilson relation,
the meaning of the overlap formula, especially
the locality properties, become clear from the point of view 
of the path-integral. 
For Dirac fermions, the overlap formalism provides a
gauge-covariant and local lattice Dirac operator
satisfying the Ginsparg-Wilson relation\cite{Ginsparg:1981bj,
Neuberger:1997fp,Kikukawa:1997qh,Neuberger:1998wv,
Hernandez:1998et}. The overlap formula was derived from the five-dimensional approach of domain wall fermion proposed by Kaplan\cite{Kaplan:1992bt}.
In the vector-like formalism of domain wall fermion\cite{Shamir:1993zy, Furman:ky,
Blum:1996jf, Blum:1997mz}, 
the local low energy effective action of the chiral mode 
precisely reproduces the overlap Dirac 
operator \cite{Vranas:1997da,Neuberger:1997bg, Kikukawa:1999sy}.
}
By this relation, it is possible to realize an exact chiral symmetry on the lattice\cite{Luscher:1998pq}. 
It is also possible to introduce Weyl fermions on the lattice and 
this opens the possibility
to formulate anomaly-free chiral lattice gauge theories\cite{Luscher:1998kn,Luscher:1998du,
Luscher:1999un,Luscher:1999mt,Luscher:2000hn,Suzuki:1999qw,Neuberger:2000wq,Adams:2000yi,
Suzuki:2000ii,Igarashi:2000zi,Luscher:2000zd,Kikukawa:2000kd,Kikukawa:2001mw,
Aoyama:1999hg}.
Although it is believed that the chiral gauge theory is a difficult case for numerical simulations
because the effective action induced by Weyl fermions 
has a non-zero imaginary part, still it would be interesting and even useful to 
develop a formulation of chiral lattice gauge theories by which one can work out 
fermionic observables numerically 
as the functions of link field with exact gauge invariance. 

In the case of U(1) chiral gauge theories,  
L\"uscher\cite{Luscher:1998du}  proved rigorously that
it is possible to construct the fermion path-integral measure 
which depends smoothly on the gauge field  and
fulfills the fundamental requirements such as 
locality, gauge-invariance, integrability and lattice symmetries.\footnote{The gauge-invariant construction by L\"uscher\cite{Luscher:1998du} 
based on the Ginsparg-Wilson relation 
provides a procedure to determine the phase of the overlap formula
 in a gauge-invariant manner for anomaly-free U(1) chiral gauge theories.
}
In this formulation, however, 
although the proof of the existence of the fermion measure is constructive,  the resulted formula of the fermion measure turns out to be rather complicated for the case of the finite-volume lattice. 
In particular, to take into account the requirements of locality and smoothness, 
it is based on the procedure to separate
the part definable in infinite volume and the part of the finite volume corrections. 
Therefore  it does not provide a formulation which is immediately usable for numerical applications. 
The purpose of this paper is
to present  a simple and closed expression  of the fermion measure (term) 
for  the U(1) chiral lattice gauge theories 
defined within the finite-volume lattice. 


This paper is organized as follows. 
In section~\ref{sec:U1-chiral-lattice-gauge-theory-review}, 
we review the construction of U(1) chiral lattice gauge theories based on the Ginsparg-Wilson relation and the reconstruction theorem of the fermion measure formulated 
by L\"uscher \cite{Luscher:1998du}. 
In section~\ref{sec:measure-term-in-V},  we describe 
our construction of the fermion measure term on the finite-volume lattice, 
which fulfills all the required properties for the reconstruction theorem. 
In our formulation, the Wilson line degrees of freedom of the link field (torons) are treated separately: 
we first construct the measure term for these degrees of freedom to take care of the global integrability.  The part of local counter term is then explicitly constructed with the local current associated with the cohomologically trivial part of  the gauge anomaly in finite volume. Combining these results, 
we finally obtain a closed formula of the measure term on the finite volume lattice, which is similar to the known expression of the measure term in the infinite volume with one parameter integration. 
Section~\ref{sec:discussion} is devoted to summary and discussions.

\section{U(1) chiral gauge theories on the lattice with exact gauge invariance}
\label{sec:U1-chiral-lattice-gauge-theory-review}

In this section, we review the construction of U(1) chiral lattice gauge theories with exact gauge invariance given by 
L\"uscher \cite{Luscher:1998du}. 
We consider U(1) gauge theories where the gauge field 
couples to $N$ left-handed Weyl fermions with charges ${\rm e}_\alpha$
satisfying the anomaly cancellation condition, 
\begin{equation}
\label{eq:anomaly-cancellation-condition}
\sum_{\alpha=1}^N {\rm e}_\alpha^3  = 0.
\end{equation}
We assume the four-dimensional lattice of the finite size $L$ and choose lattice units, 
\begin{equation}
\Gamma =
 \left\{x=(x_1, x_2, x_3, x_4)  \in \mathbb{Z}^4 \, \vert \, \, 0 \le x_\mu < L \,  (\mu = 1,2,3,4) \right\} , 
\end{equation}
and adopt the periodic boundary condition for both boson fields and fermion fields. 

\subsection{Gauge fields}
We adopt the  compact formulation of U(1) gauge theory on the lattice.  
U(1) gauge fields on $\Gamma$ then are represented by link fields,  $U(x,\mu) \, \in \text{U(1)}$. 
We require the so-called admissibility condition on the gauge fields:
\begin{equation}
\vert F_{\mu\nu}(x) \vert  < \epsilon \qquad {\rm for \ all} \ \ 
x,\mu,\nu, 
\end{equation}
where the field tensor $F_{\mu\nu}(x)$ is defined from the plaquette variables, 
\begin{eqnarray}
F_{\mu\nu}(x) &=& \frac{1}{i} {\rm ln} P_{\mu\nu}(x), 
\quad - \pi < F_{\mu\nu}(x) \le \pi , \\
P_{\mu\nu}(x)&=&U(x,\mu)U(x+\hat\mu,\nu)U(x+\hat\nu,\mu)^{-1}
U(x,\nu)^{-1}, 
\end{eqnarray}
and $\epsilon$ is a fix number in the range $0 < \epsilon < \pi/3$. 
This condition ensures that 
the overlap Dirac operator\cite{Neuberger:1997fp,Neuberger:1998wv} is a
smooth and local function of  the gauge field if $| {\rm e}_\alpha | \epsilon <1/30$
for all $\alpha$ \cite{Hernandez:1998et}.
The admissibility condition may be imposed dynamically by choosing the following action, 
\begin{equation}
S_G = \frac{1}{4g_0^2} \sum_{x\in \Gamma} \sum_{\mu,\nu}  L_{\mu\nu}(x), 
\end{equation}
where 
\begin{equation}
L_{\mu\nu}(x)= \left\{ \begin{array}{ll}
\left[ F_{\mu\nu}(x) \right]^2 
  \left\{ 1 -  \left[ F_{\mu\nu}(x) \right]^2 / \epsilon^2 \right\}^{-1}  
  & \quad \text{if } \vert F_{\mu\nu}(x) \vert < \epsilon , \\
  \infty & \quad \text{otherwise} .
  \end{array}
  \right.
\end{equation}

The admissible U(1) gauge fields can be classified by the magnetic fluxes, 
\begin{equation}
m_{\mu\nu} 
=
\frac{1}{2\pi}\sum_{s,t=0}^{L-1}F_{\mu\nu}(x+s\hat\mu+t\hat\nu) , 
\end{equation}
which are integers independent of $x$.  We denote the space of the admissible gauge fields with a given magnetic flux $m_{\mu\nu}$ by $\mathfrak{U}[m]$. As a reference point in the given 
topological sector $\mathfrak{U}[m]$, one may introduce the gauge field 
which has the constant field tensor equal to $2\pi m_{\mu\nu}/L^2 (< \epsilon)$ by
\begin{eqnarray}
V_{[m]}(x,\mu) 
&=&{\text{e}}^{-\frac{2\pi i}{L^2}\left[
L \delta_{ \tilde x_\mu,L-1} \sum_{\nu > \mu} m_{\mu\nu}
 \tilde x_\nu +\sum_{\nu < \mu} m_{\mu\nu} \tilde x_\nu
\right]}    \quad (\tilde x_\mu = x_\mu \text{ mod } L ).  
\end{eqnarray}
Then any admissible U(1) gauge field in $\mathfrak{U}[m]$ 
may be expressed as 
\begin{equation}
\label{eq:U-tilde}
U(x,\mu)=\tilde U(x,\mu) \, V_{[m]}(x,\mu) , 
\end{equation}
where  $\tilde U(x,\mu)$ stands for the dynamical degrees of freedom.
Accordingly, any local variation of the link field $U(x,\mu) \in \mathfrak{U}[m]$ should refer to 
$\tilde U(x,\mu)$:
\begin{equation}
\delta U(x,\mu) = \big\{ \delta \tilde U(x,\mu) \big\} \, V_{[m]}(x,\mu) .
\end{equation}

U(1) gauge fields on $\Gamma$ with the periodic boundary condition 
may be represented through periodic link fields on the infinite lattice: 
\begin{eqnarray}
&& U(x,\mu) \, \in \text{U(1)},    \qquad  x  \in \mathbb{Z}^4,  \\
&& U(x+L\hat \nu, \mu) = U(x,\mu) \quad \text{for all }  \mu,\nu.  
\end{eqnarray}

\subsection{Weyl fields}
Weyl fermions are introduced  based on the Ginsparg-Wilson relation.
We first consider  Dirac fields $\psi(x)$ 
which carry a Dirac index and a flavor index $\alpha = 1, \cdots, N$.  Each component $\psi_\alpha(x)$ couples to the link field,  $U(x,\mu)^{{\rm e}_\alpha}$. 
We assume that the lattice Dirac operator acting on $\psi(x)$ satisfies the Ginsparg-Wilson 
relation\footnote{In this paper, we adopt the normalization of the lattice Dirac operator so that the factor $2$ appears in the right-hand-side of the Ginsparg-Wilson relation: 
$ \gamma_5 D_L + D_L \gamma_5 = 2 D_L \gamma_5 D_L $. }, 
\begin{equation}
\label{eq:GW-rel}
\gamma_5 D_L + D_L \hat \gamma_5  = 0, \qquad \hat \gamma_5 \equiv \gamma_5(1- 2 D_L ) , 
\end{equation}
and we define the projection operators as 
\begin{equation}
\hat P_{\pm} = \left( \frac{1\pm \hat \gamma_5}{2} \right) , \quad
 P_{\pm} = \left( \frac{1\pm  \gamma_5}{2} \right) . 
 \end{equation}
The left-handed Weyl fermions, for example,  can be defined by imposing the
constraints, 
\begin{equation}
\psi_{-}(x) = \hat P_{-} \psi(x) ,  \quad
\bar \psi_-(x) = \bar \psi(x) P_{+} .   
\end{equation}
The action of the left-handed Weyl fermions is then given by 
\begin{equation}
S_W = \sum_{x\in\Gamma} \bar \psi_- (x) D_L  \psi_-(x) . 
\end{equation}

The kernel of the lattice Dirac operator in finite volume, $D_L$,  may be represented 
through the kernel of the lattice Dirac operator in infinite volume, $D$, as follows:
\begin{equation}
\label{eq:DL-in-D-infity}
D_L(x,y) = D(x,y) + \sum_{n \in \mathbb{Z}^4, n \not = 0} D(x,y+n L) ,
\end{equation}
where $D(x,y)$ is defined with a periodic link field in  infinite volume. We assume that 
$D(x,y)$ posseses the locality property given by 
\begin{equation}
\label{eq:locality-D}
\| D(x,y) \|  \le C ( 1+\| x-y \|^p ) \, {\rm e}^{-\| x-y \| / \varrho}
\end{equation}
for some constants $\varrho > 0$, $C >0$ , $p \ge 0$, 
where  $\varrho$ is the localization range of the lattice Dirac operator. 

\subsection{Path-integral measure of Weyl fermions}
\label{subsec:properties_of_measure_term}

The path-integral measure of the Weyl fermions may be defined by
the Grassmann integrations, 
\begin{equation}
{\cal D}[\psi_-] {\cal D}[\bar \psi_-]  = \prod_j d c_j  \prod_k d \bar c_k  ,   
\end{equation}
where $\{ c_j \}$ and $\{ \bar c_k \}$ are the grassman coefficients
in the expansion of the Weyl fields, 
\begin{equation}
\psi_-(x) = \sum_j  v_j(x) c_j , \quad  \bar \psi_-(x) = \sum_k \bar c_k \bar v_k(x) 
\end{equation}
 in terms of the chiral (orthonormal) basis defined by 
\begin{equation}
\hat P_{-} v_j(x) = v_j(x) ,    \quad 
\bar v_k (x) P_{+}  = \bar v_k (x) .  
\end{equation}
Since the projection operator $\hat P_{-} $ depends on the gauge field through $D$, 
the fermion measure also depends on the gauge field.  In this gauge-field dependence
of the fermion measure, there is  an ambiguity by a pure phase factor, because
any unitary transformation of the basis, 
\begin{equation}
\tilde v_j(x) = \sum_l v_l (x) \left(  {\cal Q}^{-1} \right)_{lj},  \qquad 
\tilde c_j = \sum_l   {\cal Q}_{jl}  c_l , 
\end{equation}
induces a change of the measure by the pure phase factor $\det {\cal Q}$.
This ambiguity should be fixed
so that
it fulfills the fundamental requirements such as 
locality, 
gauge-invariance, integrability and lattice symmetries. 


\subsection{Reconstruction theorem of the Weyl fermion measure}
The properties of the fermion measure can be characterized by the so-called measure term
which is given in terms of the chiral basis and its variation with respect to the gauge field,
$\delta_\eta U(x,\mu) = i \eta_\mu(x) U(x,\mu)$, 
as 
\begin{equation}
\label{eq:measure-term}
\mathfrak{L}_\eta = i \sum_j  ( v_j , \delta_\eta v_j ) . 
\end{equation}
The reconstruction theorem given in \cite{Luscher:1998du} asserts that 
if there exists a local current $j_\mu(x)$ which satisfies the following four 
properties, 
it is possible to reconstruct  the 
fermion measure (the basis $\{ v_j(x) \}$) which depends smoothly on the gauge field  and
fulfills the fundamental requirements such as 
locality\footnote{We adopt the generalized notion of locality on the lattice given in \cite{Hernandez:1998et, Luscher:1998kn, Luscher:1998du} 
for Dirac operators and 
composite fields.  See also \cite{Kadoh:2003ii} for the case of the finite volume lattice.}, 
gauge-invariance, integrability and lattice symmetries\footnote{The lattice symmetries mean
translations, rotations, reflections and charge conjugation.}: 

\vspace{1em}
\noindent{\bf Theorem} \ \  {\sl Suppose $j_\mu(x)$ is a given current with the following 
properties\footnote{Throughout this paper, 
${\text Tr}\{\cdots\}$ stands for the trace over the lattice index $x$, 
the flavor index $\alpha (=1,\cdots,N)$ and the spinor index, while  
${\text tr}$ stands for the trace over the flavor and spinor indices only. 
${\text Tr}_L\{\cdots\}$ stands for the trace over the finite lattice, $x \in \Gamma$.}:

\begin{enumerate}
\item $j_\mu(x)$  is defined for all admissible gauge fields
and depends smoothly on the link variables.

\item $j_\mu(x)$ is gauge-invariant and transforms as an axial vector
  current under the lattice symmetries. 

\item The linear functional $\mathfrak{L}_\eta= \sum_{x\in \Gamma} \eta_\mu(x) j_\mu(x)$
is a solution of the integrability condition
\begin{equation}
\label{eq:integrability-condition}
\delta_\eta \mathfrak{L}_\zeta - \delta_\zeta \mathfrak{L}_\eta
= i {\rm Tr}_L \left\{ \hat P_- [ \delta_\eta \hat P_-, \delta_\zeta
\hat P_- ] \right\} 
\end{equation}
for all periodic variations $\eta_\mu(x)$ and $\zeta_\mu(x)$.

\item The anomalous conservation law holds: 
\begin{equation}
\label{eq:anomalous-conservation-law}
\partial_\mu^\ast j_\mu(x) 
=   {\rm tr}\{ Q \gamma_5(1-D_L)(x,x) \}, \qquad
Q=\text{diag}({\rm e}_1, \cdots, {\rm e}_N) . 
\end{equation}
\end{enumerate}
Then there exists a smooth fermion integration measure in the vacuum sector such that
the associated current coincides with $j_\mu(x)$. The same is true in all other sectors
if the number of fermion flavors with $\vert {\rm e}_\alpha \vert = {\rm e} $ is even for all odd $ {\rm e} $. 
In each case the measure is uniquely determined up to a constant phase factor. 
}

\vspace{1em}
A comment is in order about the topological aspects of the reconstrtuction theorem. 
As discussed in \cite{Luscher:1998du}, it is possible to associate a U(1) bundle with the 
fermion measure.
In this point of view, 
the measure term,  $\mathfrak{L}_\eta$ defined by eq.~(\ref{eq:measure-term}), 
can be regarded as the connection of the U(1) bundle,  and 
the quantity which appears in the r.h.s. of the  integrability condition 
eq.~(\ref{eq:integrability-condition}),  \begin{equation}
\mathfrak{C}_{\eta\zeta} \equiv 
 i {\rm Tr}_L \left\{ \hat P_- [ \delta_\eta \hat P_-, \delta_\zeta
\hat P_- ] \right\}  , 
\end{equation}
is nothing but the curvature of the connection, 
\begin{equation}
\mathfrak{C}_{\eta\zeta} = \delta_\eta \mathfrak{L}_\zeta -  \delta_\zeta \mathfrak{L}_\eta . 
\end{equation} 
It is known that the integration of the curvature of a U(1) bundle 
over any two-dimensional closed surface in the base manifold
takes value of  the multiples of $2\pi$. If one parametrize a two-dimensional closed surface 
in the space of the admissible U(1) gauge fields 
by $s, t \in [ 0,2\pi]$, then one has  
\begin{equation}
\int_0^{2\pi} ds \, \int_0^{2\pi} d t \,\, 
 i {\rm Tr} \left\{ \hat P_- [ \partial_s  \hat P_-, \partial_t 
\hat P_- ] \right\}  = 2 \pi \times \text{integer} . 
\end{equation}
If (and only if) the U(1) bundle is trivial,  
these integrals of the curvature vanishes identically. 
The integrability condition eq.~(\ref{eq:integrability-condition}) asserts that it is indeed the case and 
the  fermion measure is then smooth.  

\subsection{Constructive proof of the existence of the measure term} 

In \cite{Luscher:1998du}, it is proved constructively that there exists 
a {\it local} current $j_\mu(x)$ which satisfies the properties required in the reconstruction theorem.  
In fact, the construction of the current  is not straightforward by two reasons. The first reason is that 
the measure term must be smooth w.r.t. the gauge field, but 
the topology of the space of the admissible gauge fields in finite volume is not trivial. 
The second reason is that the locality property of the current must be maintained even in finite volume.   
To take these points into account, the construction in \cite{Luscher:1998du} is made in two steps
by sperating the part definable in infinite volume from the part of the finite volume corrections. 

The procedure to sperate the part definable in infinite volume from the part of the finite volume corrections is as follows. 
As 
eq.~(\ref{eq:DL-in-D-infity}), 
one may represent the kernel of the chiral projector in finite volume $\hat P_-(x,y)$ through 
the kernel of chiral projector in infinite volume, 
$P(x,y) = \frac{1}{2}(1-\gamma_5)\delta_{xy} +  \frac{1}{2}\gamma_5 D(x,y)$,  as
\begin{equation}
\hat P_-(x,y) = \sum_{n \in \mathbb{Z}^4} P(x,y+n L) . 
\end{equation}
One may also  introduce the projector $Q_\Gamma$ acting on the fields in infinite volume as
\begin{equation}
Q_\Gamma \, \psi(x) = \left\{ 
\begin{array}{ll} \psi(x) & \text{if $x \in \Gamma, $} \\ 0 & \text{otherwise. } \end{array} \right.
\end{equation}
Using these, the right-hand-sides of the integrability condition  eq.~(\ref{eq:integrability-condition}) 
and the anomalous conservation law eq.~(\ref{eq:anomalous-conservation-law}) 
may be rewritten into 
\begin{eqnarray}
\label{eq:curvature-infinite-finite}
i {\rm Tr}_L \left\{ \hat P_- [ \delta_\eta \hat P_-, \delta_\zeta
\hat P_- ] \right\} 
&=& i {\rm Tr} \left\{ Q_\Gamma \hat P_- [ \delta_\eta \hat P_-, \delta_\zeta
\hat P_- ] \right\} 
+ \mathfrak{R}_{\eta \zeta},  
\end{eqnarray}
and
\begin{eqnarray}
 {\rm tr}\{ Q \gamma_5(1-D_L)(x,x) \} &=&  {\rm tr}\{ Q \gamma_5(1-D)(x,x) \} + r(x), 
\end{eqnarray}
respectively, where  $\mathfrak{R}_{\eta \zeta}$  and $r(x)$ are finite-volume corrections, 
\begin{eqnarray}
&& \mathfrak{R}_{\eta \zeta} = 
i \sum_{x\in\Gamma}  \sum_{y,z \in\mathbb{Z}^4} \sum_{n \in \mathbb{Z}^4, n \not = 0} 
{\rm tr}\left\{
P(x,y) \right. \nonumber\\
&&\qquad\qquad
\times \left. \left[
\delta_\eta P(y,z) \delta_\zeta P(z,x+Ln)- \delta_\zeta P(y,z) \delta_\eta P(z,x+Ln)
\right]\right\},  \\
&&\nonumber\\
&&r(x) = \sum_{n \in \mathbb{Z}^4, n\not = 0}  {\rm tr}\{ Q \gamma_5(1-D)(x,x+Ln) \} , 
\end{eqnarray}
satisfying 
\begin{eqnarray}
\label{eq:bound-R}
&& \left| \mathfrak{R}_{\eta \zeta} \right|  \le 
\kappa_1 L^{\nu_1} {\rm e}^{-L/\varrho} \| \eta \|_\infty \| \zeta \|_\infty , \\
&& \left| r(x) \right| \le  C_1  \, {\rm e}^{-L/\varrho} 
\end{eqnarray}
for some constants $\kappa_1 >0$, $\nu_1 \ge 0$ and $C_1 > 0$.
These bounds follow from the locality property of the lattice Dirac operator $D$ 
in infinite volume eq.~(\ref{eq:locality-D}). 

Then, as the first step, one constructs a local current $j_\mu^\star(x)$ in infinite volume 
so that  
1) it depends smoothly on the link variables,  2) it is gauge-invariant and transforms as an axial vector current under the lattice symmetries,  
3) 
the linear functional defined with a periodic link variables, 
\begin{equation}
\label{eq:measure-term-current-infinite-volume}
\mathfrak{K}_\eta=\sum_{x\in\Gamma} \eta_\mu^{}(x) j_\mu^\star(x), 
\end{equation}
is a solution of the integrability condition
\begin{equation}
\label{eq:integrability-condition-infinite-volume1}
\delta_\eta \mathfrak{K}_\zeta - \delta_\zeta \mathfrak{K}_\eta
= i {\rm Tr} \left\{ Q_\Gamma \hat P_- [ \delta_\eta \hat P_-, \delta_\zeta
\hat P_- ] \right\} 
\end{equation}
for all periodic  variations $\eta_\mu(x)$ and $\zeta_\mu(x)$, 
and 4) 
it satisfies the anomalous conservation law in infinite volume,
\begin{equation}
\label{eq:anomalous-conservation-law-infinite-volume1}
\partial_\mu j_\mu^\star(x) 
=   {\rm tr}\{ Q \gamma_5(1-D)(x,x) \} . 
\end{equation}
%
%
As the second step, one constructs the finite-volume correction to $\mathfrak{K}_\eta$, 
\begin{equation}
 \mathfrak{S}_\eta= \sum_{x \in \Gamma} \eta_\mu(x) \Delta j_\mu(x)
\end{equation}
with the property
\begin{equation}
| \Delta j_\mu(x) | \le \kappa_2 L^{\nu_2}  {\rm e}^{-L/\varrho} 
\end{equation}
for some constants $\kappa_2 >0$, $\nu_2 \ge 0$, so that it satisfies the conditions 1) and 2) above 
and 
\begin{equation}
\label{eq:finite-volume-correction-conditions}
\delta_\eta  \mathfrak{S}_\zeta -  \delta_\zeta \mathfrak{S}_\eta = \mathfrak{R}_{\eta\zeta} , \qquad
\partial_\mu \Delta j_\mu(x) = r(x) . 
\end{equation}
The linear functional 
$\mathfrak{L}_\eta \equiv \mathfrak{K}_\eta+ \mathfrak{S}_\eta$ then
fulfills all the required properties for the measure term {\em on the finite-volume lattice}.\footnote{In fact, the second condition in eq.~(\ref{eq:finite-volume-correction-conditions}) follows from the 
first condition and the gauge invariance of $\Delta j_\mu(x)$ \cite{Luscher:1998du}.}

\subsubsection{First step in infinite volume: locality}
\label{sec:first-step-locality}
In the first step, the explicit expression of the local current $j_\mu^\star(x)$  
is obtained \cite{Luscher:1998du}. This  is based on the two facts which hold true in infinite volume.  

The first fact is about
the  gauge anomaly associated with the Weyl fermions in the U(1) chiral lattice gauge theories, 
\begin{equation}
q(x)={\rm tr}\left\{ Q 
\gamma_5(1- D)(x,x) \right\} \qquad (x \in \mathbb{Z}^4),  
\end{equation}
which is topological by virtue of the 
Ginsparg-Wilson relation\cite{Luscher:1998pq,Kikukawa:1998pd,Fujikawa:1998if,
Adams:1998eg,Suzuki:1998yz,Chiu:1998xf}:

\vspace{1em}
\noindent{\bf Lemma 2.a} \ \  {\sl The U(1) gauge anomaly $q(x)$ has the following form:
\begin{eqnarray}
\label{eq:-cohomology-result-infinite-volume}
q(x) =  \gamma \,   \Big(\sum_\alpha e_\alpha^3\Big) 
\, \epsilon_{\mu\nu\lambda\rho}F_{\mu\nu}(x) F_{\lambda\rho}(x+\hat\mu+\hat\nu)
+ \partial_\mu^\ast  \bar k_\mu(x) ,  
\end{eqnarray} 
where $\gamma$ is a constant and $\bar k_\mu(x)$ is a local, gauge-invariant  current, which
can be constructed so that  it transforms as the axial vector current under the lattice 
symmetries.  For the anomaly-free multiple, the cohomologically non-trivial part of the gauge anomaly cancels exactly at a finite lattice spacing and the total gauge anomaly is cohomologically trivial:
\begin{equation}
q(x) 
= \partial_\mu^\ast \bar k_\mu(x) . 
\end{equation}
}

\vspace{1em}
\noindent
This result was shown in \cite{Luscher:1998kn,Fujiwara:1999fi,Fujiwara:1999fj}. 
$\gamma$ is a constant which takes the value $\gamma = \frac{1}{32 \pi^2}$ for the overlap Dirac operator \cite{Kikukawa:1998pd}.  

The second fact is about the representation of 
admissible link fields in terms of  vector potentials with the desired locality property:

\vspace{1em}
\noindent{\bf Lemma 2.b} \ \  {\sl Suppose $U(x,\mu)$ is an admissible gauge field on the infinite lattice. Then there exists a vector potential $A_\mu(x)$ such that 
\begin{eqnarray}
&& U(x,\mu) = {\rm e}^{i A_\mu(x) } ,  \quad   | A_\mu(x) |  \le \pi( 1 + 4 \| x \| ) , \\
&& F_{\mu\nu}(x) = \partial_\mu A_\nu(x)- \partial_\nu A_\mu(x) . 
\end{eqnarray}
Moreover, any other field with these properties is equal to $A_\mu(x)+\partial_\mu \omega(x)$, where the gauge function $\omega(x)$ takes values that are integer multiples of $2\pi$.
}

\vspace{1em}
\noindent
An important property of this mapping is that
the locality properties of the gauge invariant fields are 
the same independently of whether they are considered
to be functions of the link variables or the vector potential. 
To see this, let us first consider a local field which is composed from the link variables $U(x,\mu)$. 
Since the mapping 
$A_\mu(x) \rightarrow U(x,\mu)= {\rm e}^{i A_\mu(x)}$
is manifestly local,  this function is local with respect to the vector potential. 
In the other direction, let us assume a gauge invariant local field $\phi(y)$ depending on the vector potential $A_\mu(x)$. Then we remind that it is free 
to change the gauge in  constructing $A_\mu(x)$. 
In particular, we may impose a complete axial gauge taking the point $y$ as the origin. 
Around $y$ the vector potential 
$A_\mu(x)$ is locally constructed from the given link field $U(x,\mu)$. 
Thus $\phi(y)$ maps to a local function of the link variables $U(x,\mu)$ residing there. 


%


Then, for a given admissible link field $U(x,\mu)={\rm e}^{ i A_\mu(x) }
$ and any variational parameter $\eta_\mu(x)$ of compact support, 
one may define a linear functional $\mathfrak{L}_\eta^\star = 
\sum_{x \in \mathbb{Z}^4 } \eta^{}_\mu(x) j^\star _\mu(x)$ by the formula, 
\begin{eqnarray}
\label{eq:measure-term-in-infinite-volume}
\mathfrak{L}_\eta^\star 
&=& i \int_0^1 ds \, \text{Tr} \left\{ \hat P_- \big[ \partial_s \hat P_-, \delta_\eta \hat P_- \big] \right\} + \nonumber \\
&& \int_0^1 ds \, \sum_{x \in \mathbb{Z}^4 }\big\{   \eta_\mu(x)  
\bar k_\mu(x) + A_\mu(x) \delta_\eta \bar k_\mu(x) \big\} , 
\end{eqnarray}
where the differentiation and the integration with respect to  the parameter $s$ should be performed
along the one-parameter family of the admissible link fields defined by 
$U_s(x,\mu) = {\rm e}^{i  s A_\mu(x) } $. 
This linear functional $\mathfrak{L}_\eta^\star $ satisfies 
all the properties required to the measure term in infinite volume. 
In particular, the current $j^\star _\mu(x)$ is a local functional of the link variables. 
To see this property, we first note that 
it is local with respect to the vector potential $A_\mu(x)$
because of the locality properties of the kernel of the projection operator $\hat P_-(x,y)$ and  
the current $\bar k_\mu(x)$.  
We next note that 
$j^\star _\mu(x)$ is invariant 
under the gauge transformations $A_\mu(x) \rightarrow A_\mu(x) + \partial_\mu \omega(x)$
for arbitrary  gauge functions  $\omega(x)$ that are  polynomially bounded at infinity. 
Namely, taking the gauge covariance of $\hat P_-(x,y)$ and the gauge invariance of $\hat k_\mu(x)$ into account,  the change of  $\mathfrak{L}_\eta^\star$ is evaluated as  
\begin{eqnarray}
&& \int_0^1 ds \, {\rm Tr} \big\{ \hat P_- \big[  [ \omega Q, \hat P_- ], \delta_\eta \hat P_-  \big] \big\}
+ \int_0^1 ds \, \sum_{x \in \mathbb{Z}^4} \partial_\mu \omega(x) \, \delta_\eta \bar k_\mu(x)  \nonumber\\
&=& - \int_0^1 ds \, {\rm Tr} \big\{ \omega Q \, \delta_\eta \hat P_- \big\} 
+ \int_0^1 ds \, \sum_{x \in \mathbb{Z}^4} \partial_\mu \omega(x) \, \delta_\eta \bar k_\mu(x)  \nonumber\\
&=& \int_0^1 ds \sum_{x \in \mathbb{Z}^4} \omega(x) \, \delta_\eta\left\{ 
-{\rm tr}\{ Q \gamma_5 D \}(x,x) - \partial^\ast \bar k_\mu(x)  \right\}  = 0 , 
\end{eqnarray}
where the identity $\hat P_- \delta_\eta \hat P_- \hat P_- =0$ has been used.  Then,  we can regard 
$j^\star _\mu(x)$  as a local functional with respect to the link variables.

\subsubsection{Second step  in  finite volume: smoothness}

In the second step constructing the finite-volume correction $\mathfrak{S}_\eta$ which must be 
smooth with respect to the link variables, one needs to know the topological structure of the space of the admissible U(1) gauge fields in finite volume. 
It turns out that
the space $\mathfrak{U}[m]$ is isomorphic to a multi-dimensional torus times a contractible space. 
Namely, 
\begin{equation}
\label{eq:topology-Um}
\mathfrak{U}[m] \cong U(1)^4 \times \mathfrak{G}_0 \times \mathfrak{A}[m] , 
\end{equation}
where 
$\mathfrak{G}_0$ is the subset of the gauge transformations $\Lambda(x) \in U(1)$ 
satisfying $\Lambda(x)=1$ at $x=0 \text{ mod } L$,  
$\mathfrak{A}[m]$ is the space of the transverse vector potential $A_\mu^T(x)$  satisfying 
\begin{eqnarray}
&&\partial_\mu^\ast  A^T_\mu(x) = 0, \qquad
     \sum_{x\in \Gamma} A^T_\mu(x) = 0, \\
&& 
\left\vert 
\partial_\mu A^T_\nu(x)-\partial_\nu A^T_\mu(x) + 2\pi m_{\mu\nu}/L^2 
\right\vert 
< \epsilon , 
\end{eqnarray}
and $U(1)^4$ comes from the degrees of freedom of the Wilson lines. 
In fact, the following lemma provides a unique representation of 
$U(x,\mu)$  and establishes the isomorphism 
eq.~(\ref{eq:topology-Um}) \cite{Luscher:1998du}: 

\vspace{1em}
\noindent{\bf Lemma 2.c} \ \  {\sl The gauge fields $U(x,\mu)$ in the sector $\mathfrak{U}[m]$ 
are of the form
\begin{equation}
\label{eq:U-in-AT}
U(x,\mu) =V_{[m]}(x,\mu) \, \,  {\rm e}^{i A^T_\mu(x)} \, \, U_{[w]}(x,\mu) 
 \, \Lambda(x)  \, \Lambda(x+\hat\mu)^{-1}, 
\end{equation}
where 
$A^T_\mu(x)$ is the transverse vector potential  in $\mathfrak{A}[m]$ satisfying 
\begin{eqnarray}
&& 
\partial_\mu A^T_\nu(x)-\partial_\nu A^T_\mu(x) + 2\pi m_{\mu\nu}/L^2 
= F_{\mu\nu}(x) ,  
\end{eqnarray}
$U_{[w]}(x,\mu)$ represents the degrees of freedom of  the Wilson lines, 
\begin{equation}
\label{eq:link-field-wilson-lines}
U_{[w]}(x,\mu) = \left\{ 
\begin{array}{cl}
w_\mu
&\quad  \text{if $x_\mu = L-1$},  \\
1           &\quad \text{otherwise,  }  
\end{array} \right.
\end{equation}
with the phase factor $w_\mu \in U(1)$ and 
$\Lambda(x)$ is the gauge function  
in $\mathfrak{G}_0$ statisfying $\Lambda(0)=1$. 
}

\vspace{1em}
\noindent

Once the topology of the space $\mathfrak{U}[m]$ is identified as a multi-dimensional torus times a constractible space, the construction of the smooth finite-volume correction $\mathfrak{S}_\eta$ is achieved based on the bound eq.~(\ref{eq:bound-R}) and the following mathematical fact:

\vspace{1em}
\noindent{\bf Lemma 2.d} \ \  {\sl Suppose $T^n$ is the $n$-dimensional torus 
parameterized through \\
$u=({\rm e}^{i t_1}, {\rm e}^{i t_2}, \cdots, {\rm e}^{i t_n} )$ and 
$\mathfrak{C}_{kl}(t)$ is a smooth periodic tensor field on $T^n$
satisfying 
\begin{equation}
\mathfrak{C}_{kl} = - \mathfrak{C}_{lk}, \qquad
\partial_k \mathfrak{C}_{lj}+\partial_l \mathfrak{C}_{jk}+\partial_j \mathfrak{C}_{kl} = 0.
\end{equation}
If the associated magnetic fluxes, 
\begin{equation}
\mathfrak{I}_{kl}=\int_0^{2\pi} dt_k dt_l \, \mathfrak{C}_{kl}, 
\end{equation}
vanish, there exists smooth periodic vector field $\mathfrak{B}_k(t)$ such that 
$\mathfrak{C}_{kl}= \partial_k \mathfrak{B}_l - \partial_l \mathfrak{B}_k$ and 
\begin{equation}
\left| \mathfrak{B}_k(t) \right| \le \pi(n-1) \text{sup}_{r,k,l} \left| \mathfrak{C}_{kl}(r) \right| . 
\end{equation}
}

\vspace{1em}
\noindent
In fact, 
$\mathfrak{R}_{\eta \zeta}$ on the multi-dimensional torus $T^n \cong U(1)^4 \times \mathfrak{G}_0$
turns out to satisfy all the premises of the above lemma and a solution of the integrability condition 
follows immediately from the lemma, which corresponds to the finite-volume correction term
$\mathfrak{S}_{\eta}$ 
for $U(x,\mu)=V_{[m]}(x,\mu) \, \, U_{[w]}(x,\mu)  \, \Lambda(x)  \, \Lambda(x+\hat\mu)^{-1}$
with the longitudinal variation $\eta_\mu^L(x)$\footnote{$G_L(z)$ is the Green function defined by 
\begin{equation}
\partial_\mu^\ast \partial_\mu^{} G_L(z) = \delta_{z,0}-L^{-4}, \quad
\sum_{z \in \Gamma} G_L(z) = 0. 
\end{equation}
}:
\begin{equation}
 \eta_\mu^L(x) = 
L^{-4} \sum_{y\in \Gamma} \eta_\nu(y)
+\sum_{y \in \Gamma} \partial_\mu G_L(x-y) \partial_\nu^\ast \eta_\nu(y) . 
\end{equation}
It is then extended to the transverse degrees of freedom by the integration along 
the one-parameter family 
\begin{equation}
U_t(x,\mu)=V_{[m]}(x,\mu) \, \,  {\rm e}^{i t A^T_\mu(x)} \, \, U_{[w]}(x,\mu) 
 \, \Lambda(x)  \, \Lambda(x+\hat\mu)^{-1} \qquad (t \in [0,1])
 \end{equation} 
 as follows:
\begin{equation}
\mathfrak{S}_{\eta}=  \mathfrak{S}_{\eta^L}\vert_{t=0} + \int_0^1 dt  \, \, 
\mathfrak{R}_{\zeta \eta}\vert_{\zeta_\mu = A_\mu^T}. 
\end{equation}
This completes the construction of the finite-volume correction term $\mathfrak{S}_{\eta}$. 

\section{A simple construction of the mesure term on the finite volume lattice}
\label{sec:measure-term-in-V}

In the original construction by L\"uscher \cite{Luscher:1998du}, 
although the proof is constructive,  the explicit formula of the measure term turns out to be complicated.  In particular, it is based on the separate
treatment of the part definable in infinite volume and the part of the finite volume corrections. 
Therefore it does not provide a formulation which is immediately usable for practical numerical applications. 

In this section, we 
describe our construction of the measure term on the finite volume lattice. 
We fisrt discuss the parametrization of the link fields in finite volume and their variations. 
We next state two useful results which hold true in finite volume: 
{\sl the gauge anomaly cancellation in finite volume} and {\sl the property of the
curvature term for the Wilson lines}.  Using these results,  
we write down a closed formula of the measure term directly within the finite volume theory. 
In our construction, the Wilson line degrees of freedom of the link field (torons) are treated 
separately to take care of the global integrability.  The part of local counter term is then explicitly constructed  in finite volume 
with the local current associated with the cohomologically trivial part of  the gauge anomaly. 
Only in the final step to establish the locality property of the measure term current, we follow the procedure to separate the part definable in infinite volume from the part of the finite volume corrections 
as in the original construction \cite{Luscher:1998du}. 

\subsection{Parametrization of the link fields and their variations in finite volume}

In our construction of the measure term in finite volume, 
we adopt the parametrization of the link fields given by eq.~(\ref{eq:U-in-AT}).  
When a link field $U(x,\mu)$ is parameterized by eq.~(\ref{eq:U-in-AT}),   the parametrization is 
unique and the each factors, $A_\mu^T(x)$, $U_{[w]}(x,\mu)$ and $\Lambda(x)$, may be regarded as 
the smooth functionals of the original link field $U(x,\mu)$.  

Accordingly, 
the variation of the link field, 
\begin{equation}
\delta_\eta U(x,\mu) = i \, \eta_\mu(x) \, U(x,\mu), 
\end{equation}
may be decomposed as follows:
\begin{equation}
\eta_\mu(x) = \eta^T_\mu(x) + \eta_{\mu [w]}(x) + \eta_\mu^\Lambda(x) .   
\end{equation}
$\eta^T_\mu(x)$ is the transverse part of $\eta_\mu(x)$ defined by
\begin{equation}
\partial_\mu^\ast \eta^T_\mu(x) = 0, \qquad \sum_{x \in \Gamma} \eta^T_\mu(x) =0, 
\end{equation}
which may be given explicitly as
\begin{equation}
\eta^T_\mu(x)  = \sum_{y\in\Gamma} G_L(x-y) 
\partial_\lambda^\ast (\partial_\lambda \eta_\mu(x) - \partial_\mu \eta_\lambda(x)).  
\end{equation}
$\eta_{\mu [w]}(x)$ is the variation along the Wilson lines defined by 
\begin{equation}
\label{eq:variation-parameter-wilson-lines}
\eta_{\mu [w]}(x) =\sum_{\nu} \eta_{(\nu)} \, \delta_{\mu\nu} \, \delta_{x_\nu,L-1} , \qquad
 \eta_{(\nu)} = L^{-3} \sum_{y\in \Gamma} \eta_\nu(y) .  
\end{equation}
$\eta_\mu^\Lambda(x)$ is the variation of the gauge degrees of freedom in the form,
\begin{equation}
\eta_\mu^\Lambda(x) = - \partial_\mu \omega_\eta(x) ,  \qquad \omega_\eta(0) = 0.
\end{equation}
This decomposition is also unique by the following reason:  for an arbitrary periodic vector field 
$\eta_\mu(x)$, the vector field defined by 
$a_\mu(x)=\eta_\mu(x)-\eta_\mu^T(x) - \eta_{ \mu [w]}(x)$ has the vanishing field tensor
$\partial_\mu a_\nu(x) - \partial_\nu a_\mu(x)=0$ and the vanishing wilson lines
$\sum_{s=0}^{L-1} a_\mu(x+s \hat \mu)=0$. 
Then, the sum $\omega_\eta(x)$ of the vector field
$a_\mu(x)$ along any lattice path from $x$ to the origin $x=0$ is independent of the
chosen path,  periodic in $x$ and $\omega_\eta(0)=0$.  It gives the gauge function which reproduces 
$a_\mu(x)$ in the pure gauge form, $a_\mu(x) = -\partial_\mu \omega_\eta(x)$.  This proves the uniqueness of the decomposition. 
The action of the differential operator $\delta_\eta$ to each factors, $A_\mu^T(x)$, 
$U_{[w]}(x,\mu)$ and $\Lambda(x)$, is then given as follows:
\begin{eqnarray}
&& \delta_\eta A_\mu^T(x) = \eta_\mu^T(x) , \\
&& \delta_\eta U_{[w]}(x,\mu) = i \, \eta_{\mu [w]}(x) \, U_{[w]}(x,\mu) , \\
&& \delta_\eta \Lambda(x)  
= i \, \omega_\eta(x) \,  \Lambda(x) . 
\end{eqnarray}

\subsection{Useful results in finite volume}

\subsubsection{Gauge anomaly cancellation}
In finite volume,  the U(1) gauge anomaly is given by the formula, 
\begin{equation}
q^{}_L(x)={\rm tr}\left\{ Q 
\gamma_5(1- D_L)(x,x) \right\} \qquad (x \in \Gamma),  
\end{equation}
which is topological \cite{Hasenfratz:1998ri,Narayanan:sk,Narayanan:ss} in the sense that
\begin{equation}
\sum_{x \in \Gamma} q^{}_L(x) = \text{integer} . 
\end{equation}
For this gauge anomaly in finite volume, it is possible to establish the similar result as 
the lemma 2.a:

\vspace{2em}
\noindent{\bf Lemma 3.a} \ \  {\sl For the anomaly-free multiplet satisfying 
the condition eq.~(\ref{eq:anomaly-cancellation-condition}), 
the U(1) gauge anomaly $q^{}_L(x)$ has the following form in sufficiently large volume $L^4$:
\begin{equation}
\label{eq:-cohomology-result-finite-volume}
q^{}_L(x) 
= \partial_\mu^\ast k_\mu(x) \qquad (x \in \Gamma), 
\end{equation}
where $k_\mu(x)$ is a local, gauge-invariant  current, which
can be constructed so that  it transforms as the axial vector current under the lattice 
symmetries.  }

\vspace{2em}
\noindent
This result  was first obtained by combining the result in the infinite 
lattice, eq.~(\ref{eq:-cohomology-result-infinite-volume}) \cite{Luscher:1998kn,Fujiwara:1999fi,Fujiwara:1999fj},  and the result of the analysis of the finite volume correction $r(x)$ \cite{Igarashi:2002zz}. Namely, 
\begin{equation}
k_\mu(x) = \bar k_\mu(x) + \Delta k_\mu(x) , 
\end{equation}
where $\Delta k_\mu(x)$ satisfies 
\begin{equation}
\label{eq:bound-kmu}
| \Delta k_\mu(x) | \le \kappa_3 L^{\nu_3}  {\rm e}^{-L/\varrho} 
\end{equation}
for some constants $\kappa_3 >0$, $\nu_3 \ge 0$ and 
\begin{equation}
r(x) = \partial^\ast_\mu   \Delta k_\mu(x) \qquad (x \in \Gamma)  
\end{equation}
in sufficiently large volume $L^4$. 
However, as shown in \cite{Kadoh:2003ii}, 
it is possible to derive the same result 
directly from the gauge anomaly in finite volume $q_L(x)$ 
without the separate treatment of $q(x)$ and $r(x)$.  This work also provides a procedure 
to work out 
the local current $k_\mu(x)$ explicitly, which can be implemented numerically \cite{Kadoh:2004uu}. 



\subsubsection{A solution of the integrability condition for the Wilson lines}
\label{subsubsec:prop-curvature-term}

The curvature terms associated with the Wilson lines
have special properties which turn out to be useful in the construction of 
a solution of the integrability condition eq.~(\ref{eq:integrability-condition}). 
Let us parametrize the Wilson lines $U_{[w]}(x,\mu) $ defined by eq.~(\ref{eq:link-field-wilson-lines}) as  
\begin{equation}
\label{eq:wilson-lines-parameter}
w_\mu = \exp( i t_\mu ),  \quad  t_\mu \in [0, 2\pi)  \quad (\mu =1,2,3,4) , 
\end{equation}
and the variational parameters in the directions of the Wilson lines as
\begin{equation}
\label{eq:wilson-lines-variation}
\lambda_{\mu(\nu)}(x) = \frac{1}{i} \, \partial_{t_\nu} U_{[w]}(x,\mu) \cdot U_{[w]}(x,\mu)^{-1}  
= \delta_{\mu\nu} \delta_{x_\nu,L-1}.    
\end{equation}
Then the curvature term for the Wilson lines reads
\begin{eqnarray}
 i {\rm Tr}_L
 \left\{ \hat P_- [ \delta_{\lambda_{(\mu)}} \hat P_-, \delta_{\lambda_{(\nu)}}
\hat P_- ] \right\} _{U=U_{[w]} V_{[m]}}
&=& 
i {\rm Tr}_L \left\{
\hat P_- [ \partial_{t_\mu} \hat P_- , \partial_{t_\nu} \hat P_- ] \right\} _{U=U_{[w]} V_{[m]}}
\nonumber\\
&\equiv&  \mathfrak{C}_{\mu\nu}(t),  \quad \qquad t= (t_1,t_2,t_3,t_4) .  
\end{eqnarray}
Then, the following lemma holds true: 

\vspace{1.5em}
\noindent{\bf Lemma 3.b} \ \ {\sl 
In anomaly-free theories,  the curvature term for the Wilson lines
$\mathfrak{C}_{\mu\nu}(t)$, 
which possesses the properties 
\begin{equation}
\label{eq:properties-of-curvature-1}
\mathfrak{C}_{\mu\nu}(t)  = - \mathfrak{C}_{\nu\mu}(t), \qquad
  \partial_\mu \mathfrak{C}_{\nu\rho}(t) 
+\partial_\nu \mathfrak{C}_{\rho\mu}(t) 
+\partial_\rho \mathfrak{C}_{\mu\nu}(t) 
= 0 , 
\end{equation}
satisfies the bound
\begin{equation}
\label{eq:properties-of-curvature-2}
\left\vert \mathfrak{C}_{\mu\nu} (t) \right\vert  \le  \kappa_4 L^{\nu_4} {\rm e}^{-L/\varrho} 
\end{equation}
for certain positive constants $\kappa_4$ and $\nu_4$. 
For a sufficiently large volume $L^4$, 
it then follows that
\begin{equation}
\label{eq:properties-of-curvature-3}
\int_0^{2\pi} d t_\mu \, \int_0^{2\pi} d t_\nu \,\, \mathfrak{C}_{\mu\nu}(t)  =0 , 
\end{equation}
and  there exists smooth periodic vector field $\mathfrak{W}_\mu(t)$ such that 
\begin{equation}
\mathfrak{C}_{\mu\nu}(t)= \partial_\mu \mathfrak{W}_\nu(t)
                                         - \partial_\nu \mathfrak{W}_\mu(t), \qquad
\left| \mathfrak{W}_\mu(t) \right| \le 3 \pi  \, \text{sup}_{t,\mu,\nu} \left| \mathfrak{C}_{\mu\nu}(t) \right| . 
\end{equation}
}


\vspace{1.5em}
The proof of this lemma  
is based on the fact that
in infinite-volume 
the periodic link field which represents the degrees of freedom of the Wilson lines
can be written in the pure-gauge form, 
\begin{equation}
\label{eq:wilon-line-in-pure-gauge}
U_{[w]}(x,\mu) = \Lambda_{[w]}(x) \Lambda_{[w]}(x+\hat \mu)^{-1} ,   \quad 
\Lambda_{[w]}(x)= \prod_\mu ( w_\mu )^{n_\mu} \quad \text{for  } x- n L \in \Gamma, 
\end{equation}
and therefore
the gauge-invariant function of the link field in infinite volume is actually independent of the 
degrees of freedom of the Wilson lines.  In fact, from eqs.~(\ref{eq:curvature-infinite-finite}) 
and (\ref{eq:bound-R}), 
$\mathfrak{C}_{\mu\nu}$ may be written as 
\begin{equation}
\mathfrak{C}_{\mu\nu}
= i {\rm Tr} \left\{ Q_\Gamma \hat P_- [ \delta_{\lambda_{(\mu)}} \hat P_-, \delta_{\lambda_{(\nu)}}
\hat P_- ] \right\} 
+ \mathfrak{R}_{\lambda_{(\mu)} \lambda_{(\nu)}},  
\end{equation}
where $ \mathfrak{R}_{\lambda_{(\mu)} \lambda_{(\nu)}}$ satisfies the bound
\begin{eqnarray}
\label{eq:bound-R-Wilson-lines}
&& \left| \mathfrak{R}_{\lambda_{(\mu)} \lambda_{(\nu)}} \right|  \le 
\kappa_4 L^{\nu_4} {\rm e}^{-L/\varrho} 
\end{eqnarray}
for some constants $\kappa_4 >0$, $\nu_4 \ge 0$ and $C_4 > 0$.
We then recall the fact that 
there exists the measure term 
$\mathfrak{K}_\eta = 
\sum_{x \in \Gamma} \eta_\mu(x) j_\mu^\star(x)$ given by 
eq.~(\ref{eq:measure-term-current-infinite-volume}), 
which satisfies the integrability condition eq.~(\ref{eq:integrability-condition-infinite-volume1}). 
The current $j_\mu^\star(x)$ is defined for all admissible gauge fields in 
infinite volume and it is local and gauge-invariant.  Therefore, as discussed above,
the current $j_\mu^\star(x)$ 
is actually independent of the 
Wilson lines and the curvature of $\mathfrak{K}_\eta$ 
evaluated  in the directions of the Wilson lines
vanishes identically. Namely, 
\begin{equation}
 i {\rm Tr} \left\{ Q_\Gamma \hat P_- [ \delta_{\lambda_{(\mu)}} \hat P_-, \delta_{\lambda_{(\nu)}}
\hat P_- ] \right\} 
=\delta_{\lambda_{(\mu)}}  \mathfrak{K}_{\lambda_{(\nu)}}
-\delta_{\lambda_{(\nu)}}  \mathfrak{K}_{\lambda_{(\mu)}}
= 0 .
\end{equation}
Then one can see that
the curvature for the Wilson lines, $\mathfrak{C}_{\nu\lambda}$,
itself satisfies the bound eq.~(\ref{eq:properties-of-curvature-2})
and because of this bound, the two-dimensional integration of the curvature, 
which should be a multiple of $2\pi$, 
must vanish identically for a sufficiently large $L$.
The existence of the smooth periodic vector field $\mathfrak{W}_\mu(t)$ then follows from 
the lemma 2.d (the lemma 9.2 in \cite{Luscher:1998du}). 

The properties of the curvature term $\mathfrak{C}_{\mu\nu}$ for the Wilson lines
given by eq.~(\ref{eq:properties-of-curvature-1}) and (\ref{eq:properties-of-curvature-2})
are useful because 
it implies that $\mathfrak{C}_{\mu\nu}$ itself satisfies the premise of the lemma 2.d (the lemma 9.2 
in \cite{Luscher:1998du}) and by using the lemma, 
one can construct  a solution of the integrability condition,  
\begin{equation}
 \left.  \left\{ 
 \delta_{\lambda_{(\mu)}} {\mathfrak{W}}_{\nu}
-\delta_{\lambda_{(\nu)}} {\mathfrak{W}}_{\mu}
 \right\}\right\vert_{U = U_{[w]} V_{[m]} }
=
 \mathfrak{C}_{\mu\nu} , 
\end{equation}
from $\mathfrak{C}_{\mu\nu}$ directly. 
Explicitly, it may be given by the formulae, 

\begin{eqnarray}
\mathfrak{W}_{4}
&=& \frac{1}{2\pi} \int_0^{2\pi} d r_4  \int_0 ^{(t_1,t_2,t_3)}  
\{ dr_1 \mathfrak{C}_{14} +dr_2 \mathfrak{C}_{24} +dr_3 \mathfrak{C}_{34}  \} ,  \nonumber\\
\mathfrak{W}_{3}
&=& 
\int_0^{t_4} dr_4  \mathfrak{C}_{43}  - \frac{t_4}{2\pi} \int_0^{2\pi} dr_4  \mathfrak{C}_{43}
+
\left[ \frac{1}{2\pi} \int_0^{2\pi} d r_3  \int_0 ^{(t_1,t_2)}  
\{ dr_1 \mathfrak{C}_{13} +dr_2 \mathfrak{C}_{23} \} \right]_{t_4=0} , 
 \nonumber\\
\mathfrak{W}_{2}
&=& 
 \int_0^{t_4} dr_4  \mathfrak{C}_{42}  - \frac{t_4}{2\pi} \int_0^{2\pi} dr_4  \mathfrak{C}_{42} \nonumber\\
 &&
+\left[ \int_0^{t_3} dr_3  \mathfrak{C}_{32}  
         - \frac{t_3}{2\pi} \int_0^{2\pi} dr_3  \mathfrak{C}_{32} \right]_{t_4=0} 
+\left[ \frac{1}{2\pi} \int_0^{2\pi} d r_2  \int_0 ^{(t_1)}  \{ dr_1 \mathfrak{C}_{12} \} \right]_{t_4=t_3=0} , 
\nonumber\\
\mathfrak{W}_{1}
&=& \int_0^{t_4} dr_4  \mathfrak{C}_{41}  - \frac{t_4}{2\pi} \int_0^{2\pi} dr_4  \mathfrak{C}_{41} \nonumber\\
&&+\left[\int_0^{t_3} dr_3  \mathfrak{C}_{31}  
             - \frac{t_3}{2\pi} \int_0^{2\pi} dr_3  \mathfrak{C}_{31}\right]_{t_4=0} 
+\left[\int_0^{t_2} dr_2  \mathfrak{C}_{21}  
        - \frac{t_2}{2\pi} \int_0^{2\pi} dr_2  \mathfrak{C}_{21}\right]_{t_4=t_3=0} . \nonumber\\
        \label{eq:measure-term-Wilson-lines}
\end{eqnarray}
It follows from the properties of $\mathfrak{C}_{\mu\nu}$
that this solution is periodic and  smooth
with respect to the Wilson lines $U_{[w]} $ and 
satisfies the bound
\begin{equation}
\label{eq:measure-term-Wilson-lines-bound}
\left\vert \, {\mathfrak{W}}_{\nu}  \, \right\vert 
 \le  \kappa_5 L^{\nu_5} {\rm e}^{-L/\varrho},  
\end{equation}
for certain positive constants $\kappa_5$ and $\nu_5$. 
It also follows that this measure term is gauge invariant. 
Then 
one may  introduce the linear functional 
of the variational parameters in the directions of the Wilson lines $\eta_{\mu[w]}(x)$
at the gauge field $U(x,\mu)=U_{[w]}(x,\mu) V_{[m]}(x,\mu)$ by
\begin{equation}
{\mathfrak{W}}_\eta \vert_{U=U_{[w]} V_{[m]}, \eta=\eta_{[w]}}
= \sum_\nu \eta_{(\nu)} {\mathfrak{W}}_{\nu} . 
\end{equation}
This provides the measure term at the gauge field $U(x,\mu)=U_{[w]}(x,\mu) V_{[m]}(x,\mu)$. 


\subsection{A closed formula of the measure term}

We now construct the measure term 
for the generic admissible U(1) gauge fields in the given topological sector 
$\mathfrak{U}[m]$. 
For this purpose, 
we introduce a vector potential defined by
\begin{eqnarray}
&& \tilde A_\mu^\prime(x) = A_\mu^T(x) - \frac{1}{i}\partial_\mu \Big[ \ln \Lambda(x) \Big] ; \qquad
\frac{1}{i} \ln \Lambda(x) \in (-\pi, \pi] ,  
\end{eqnarray}
and choose a  one-parameter family of the gauge fields as 
\begin{equation}
\label{eq:family-U-s}
U_s(x,\mu)= {\rm e}^{i s \tilde A_\mu^\prime(x)} \,  U_{[w]}(x,\mu) \,   V_{[m]}(x,\mu),\qquad   0 \le s \le 1 .
\end{equation}
Then we consider the linear functional of the variational parameter $\eta_\mu(x)$,  
which is given in terms of quantities defined on the finite-volume lattice:
\begin{eqnarray}
\label{eq:measure-term-finite-volume}
\mathfrak{L}_\eta^\diamond &=& i \int_0^1 ds \, 
{\rm Tr}_L \left\{ \hat P_- [ \partial_s \hat P_-, 
\delta_\eta \hat P_- ] \right\} 
\nonumber\\
&+& 
\delta_\eta \, 
\int_0^1 ds \, \sum_{x \in \Gamma} 
 \left\{   \tilde A_\mu^\prime(x)  \,   k_\mu(x) \right\}   
+{\mathfrak{W}}_\eta \vert_{U=U_{[w]} V_{[m]}, \eta=\eta_{[w]}},
\end{eqnarray}
where $k_\mu(x)$ is the gauge-invariant local current
which satisfies 
$\partial_\mu^\ast k^{}_\mu(x) = q^{}_L(x)$ and 
transforms as an axial vector field under the lattice symmetries. 
${\mathfrak{W}}_\eta \vert_{U=U_{[w]} V_{[m]}, \eta=\eta_{[w]} }$ is the additional measure term 
at the gauge field $U=U_{[w]} V_{[m]}$
with the variational parameters in the directions of the Wilson lines
$\eta_{\mu [w]}(x)$. 
The current $j_\mu^\diamond(x)$ defined by 
eq.~(\ref{eq:measure-term-finite-volume}), 
\begin{equation}
{\mathfrak{L}}_\eta^\diamond = \sum_{x \in \Gamma} \eta_\mu(x) j_\mu^\diamond(x), 
\end{equation}
may be regarded as a functional of the link variable $U(x,\mu)$ through the dependences on
$A_\mu^T(x)$, $\Lambda(x)$ ($\ln \Lambda(x)$), $U_{[w]}(x,\mu)$ and $V_{[m]}(x,\mu)$. 
The action of the differential operator $\delta_\eta$ to 
the vector potential $\tilde A_\mu^\prime(x)$ is evaluated as
\begin{eqnarray}
\delta_\eta \tilde A_\mu^\prime(x)
&=& 
\delta_\eta  A_\mu^T(x) - \partial_\mu \left[\frac{1}{i} \{ \delta_\eta \Lambda(x) \}\, \Lambda(x)^{-1} \right] 
=\eta_\mu^T(x) - \partial_\mu \omega_\eta(x) \nonumber\\
&=&
\eta_\mu(x)-\eta_{\mu [w]}(x) , 
\end{eqnarray}
and the variation of $U_s(x,\mu)$ is given by 
\begin{equation}
\delta_\eta U_s(x,\mu) = i \left[
 s(\eta_\mu(x) -\eta_{\mu [w]}(x)  ) + \eta_{\mu [w]}(x) \right] U_s(x,\mu) . 
\end{equation}

The linear functional 
so obtained, however, does not respect the lattice symmetries. 
(The first- and second- terms in the r.h.s. of eq.~(\ref{eq:measure-term-finite-volume})
transform properly, but the third term 
does not respect the lattice symmetries.)  
In order to make it
to transform as a pseudo scalar field under the lattice 
symmetries, we should average it over the lattice symmetries with the appropriate 
weights so as to project to the pseudo scalar component. 
Namely, we take the average as follows\footnote{In doing the average, one should note the fact that 
under the lattice symmetries the Wilson lines $U_{[w]}(x,\mu)$ 
are transformed  to other Wilson lines $U_{[w^\prime]}(x,\mu) $ {\em modulo gauge transformations}, 
$\{ U_{[w]}(x,\mu)\}^{R^{-1}}  = U_{[w^\prime]}(x,\mu) \Lambda(x) \Lambda(x+\hat \mu)^{-1}$. 
Accordingly, the variational parameter $\eta_{\mu [w]}(x)$ is transformed as 
$\{\eta_{\mu [w]}(x)\}^{R^{-1}} = \eta_{\mu [w^\prime]}(x) - \partial_\mu \omega(x)$ with a certain 
periodic gauge funciton $\omega(x)$. }:
\begin{equation}
\label{eq:average-L-diamond}
\bar{\mathfrak{L}}^\diamond_{\eta} = \frac{1}{2^4 4!} \sum_{R \in  O(4,\mathbb{Z})} \det R \, \, 
 \mathfrak{L}^\diamond_{\eta}\vert_{U\rightarrow\{U\}^{R^{-1}},  
         \eta_\mu\rightarrow\{\eta_\mu \}^{R^{-1}} } . 
\end{equation}

Our main result is then stated as follows:

\vspace{1em}
\noindent{\bf Lemma 3.c} \ \ {\sl The current $j_\mu^\diamond(x)$ defined by 
eq.~(\ref{eq:measure-term-finite-volume}), 
\begin{equation*}
{\mathfrak{L}}_\eta^\diamond = \sum_{x \in \Gamma} \eta_\mu(x) j_\mu^\diamond(x), 
\end{equation*}
fulfills all the properties required for the reconstruction theorem except the transformation property 
under the lattice symmetries. It may be corrected by invoking the average 
eq.~(\ref{eq:average-L-diamond}) over 
the lattice symmetries with the appropriate 
weights so as to project to the pseudo scalar component. 
}

\vspace{1em}
\subsubsection{Proof of the lemma 3.c}

Although it is quite similar to that of theorem 5.3 in \cite{Luscher:1998du}, 
we give the proof of the lemma 3.c here for completeness.  

\begin{enumerate}
\item {\it Smoothness.}  By construction, $ j_\mu^\diamond(x)$ is defined for all admissible gauge fields. 
It depends smoothly on $\tilde A_\mu^\prime(x)$ and $U_{[w]}(x,\mu)$ 
because $\hat P_-$ and $k_\mu$ are smooth functions of $U_s(x,\mu)$. 
Although $\tilde A_\mu^\prime(x)$ is not continuous 
when $\Lambda(x) = -1$ at some points $x$ because of  the cut in $\ln \Lambda(x)$,  its discontinuity is always 
in the pure-gauge form 
\begin{equation}
\text{disc.} \{ \tilde A_\mu^\prime(x) \} = - \partial_\mu \omega(x) ; \qquad \omega(0)  = 0,  
\end{equation}
where the gauge function $\omega(x)$ takes values that are integer multiples of $2\pi$.
Then, any smooth functionals of $\tilde A_\mu^\prime(x)$ are smooth with respect to the 
link field $U(x,\mu)$, if they are gauge-invariant under the gauge transformations 
$\tilde A^\prime_\mu(x) \rightarrow \tilde A^\prime_\mu(x) + \partial_\mu \omega(x)$ 
for arbitrary  periodic gauge functions $\omega(x)$ satisfying $\omega(0)=0$.
The current $j_\mu^\diamond(x)$ is indeed gauge-invariant under such gauge transformations. 
Namely, 
taking the gauge covariance of $\hat P_-(x,y)$ and the gauge invariance of $k_\mu(x)$ into account,  the change of  $\mathfrak{L}_\eta^\diamond$ under the gauge transformations 
is evaluated as  
\begin{eqnarray}
&& \int_0^1 ds \, {\rm Tr} \big\{ \hat P_- \big[  [ \omega Q, \hat P_- ], \delta_\eta \hat P_-  \big] \big\}
+ \int_0^1 ds \, \sum_{x \in \Gamma} \partial_\mu \omega(x) \, \delta_\eta k_\mu(x)  \nonumber\\
&=& - \int_0^1 ds \, {\rm Tr} \big\{ \omega Q \, \delta_\eta \hat P_- \big\} 
+ \int_0^1 ds \, \sum_{x \in \Gamma} \partial_\mu \omega(x) \, \delta_\eta k_\mu(x)  \nonumber\\
&=& \int_0^1 ds \sum_{x \in \Gamma} \omega(x) \, \delta_\eta\left\{ 
-{\rm tr}\{ Q \gamma_5 D_L \}(x,x) - \partial_\mu^\ast k_\mu(x)  \right\}  = 0 , 
\end{eqnarray}
where the identity $\hat P_- \delta_\eta \hat P_- \hat P_- =0$ has been used.  
%

\item {\it Gauge invariance and symmetry properties.} 
The gauge invariance of $ j_\mu^\diamond(x)$ has been shown above.  
The transformation properties of $j_\mu^\diamond(x)$ under the lattice symmetries are also evident 
from the average eq.~(\ref{eq:average-L-diamond}). 
%

\item {\it Integrability condition.}  From the definition of $\mathfrak{L}_\eta^\diamond$, 
eq.~(\ref{eq:measure-term-finite-volume}), one finds immediately that the second term does not contribute the curvature 
$\delta_\eta \mathfrak{L}_\zeta^\diamond-\delta_\zeta \mathfrak{L}_\eta^\diamond$ 
and the third term gives the curevature term at the Wilson lines,  $U=U_{[w]} V_{[m]}$, with 
the variational parameters $\eta, \zeta=\eta_{[w]}$.  
Taking the identity
${\rm Tr}_L \left\{ \delta_1 \hat P_- \delta_2 \hat P_- \delta_3\hat P_- \right\} =0$ into account, the curvature is evaluated as 
\begin{eqnarray}
\delta_\eta \mathfrak{L}_\zeta^\diamond-\delta_\zeta \mathfrak{L}_\eta^\diamond
&=& 
 i \int_0^1 ds \, 
{\rm Tr} \left\{ 
  \hat P_- [ \delta_\eta \partial_s \hat P_-, \delta_\zeta \hat P_- ] 
-\hat P_- [ \delta_\zeta \partial_s \hat P_-, \delta_\eta \hat P_- ] 
\right\} \nonumber\\
&&\quad\qquad\qquad + \left. 
 i {\rm Tr} \left\{ \hat P_- [ \delta_\eta \hat P_-, 
\delta_\zeta \hat P_- ] \right\} 
 \right\vert_{U = U_{[w]} V_{[m]}; \eta, \zeta=\eta_{[w]}} \nonumber\\
 &=&
 i \int_0^1 ds \,  \partial_s {\rm Tr} \left\{ 
  \hat P_- [ \delta_\eta  \hat P_-, \delta_\zeta \hat P_- ] 
\right\} \nonumber\\
&&\quad\qquad\qquad + \left. 
 i {\rm Tr} \left\{ \hat P_- [ \delta_\eta \hat P_-, 
\delta_\zeta \hat P_- ] \right\} 
 \right\vert_{U = U_{[w]} V_{[m]}; \eta, \zeta=\eta_{[w]}}. 
\end{eqnarray}
After the integration in the first term, 
the contribution from the  lower end of the integration range 
exactly cancels with the second term
because the variational parameters in this contribution
is restricted to $\eta_{\mu [w]}(x)$: 
\begin{equation}
\delta_\eta U_s(x,\mu)  \, U_s(x,\mu)^{-1}  \vert_{s=0} 
= [ s(\eta_\mu(x)- \eta_{\mu [w]}(x) ) + \eta_{\mu [w]}(x) ]_{s=0} = \eta_{\mu [w]}(x). 
\end{equation}


\item {\it Anomalous conservation law.}
If one sets $\eta_\mu(x) = - \partial_\mu \omega(x)$ (where $\omega(x)$ is any lattice function on $\Gamma$ with $\omega(0)=0$), the left-hand side of eq.~(\ref{eq:measure-term-finite-volume}) becomes
\begin{equation}
\sum_{x \in \Gamma} \omega(x) \, \partial_\mu^\ast  j_\mu^\diamond(x) . 
\end{equation}
On the other hand, using the identities
\begin{equation}
\delta_\eta \hat P_- = i s \left[ \omega Q, \hat P_- \right] , \qquad
\delta_\eta k_\mu(x) = 0, 
\end{equation}
the right-hand side is evaluated as 
\begin{eqnarray}
&& - \int_0^1 ds \, s \,  {\rm Tr} \big\{ \omega Q \, \partial_s \hat P_- \big\} 
- \int_0^1 ds \, \sum_{x \in \Gamma} \partial_\mu \omega(x) \,  k_\mu(x)
\nonumber\\
&=& 
-\sum_{x \in \Gamma} \omega(x) \,{\rm tr}\{ Q \gamma_5 D_L \}(x,x)
+
\int_0^1 ds \sum_{x \in \Gamma} \omega(x) \, \left\{ 
{\rm tr}\{ Q \gamma_5 D_L \}(x,x) + \partial_\mu^\ast k_\mu(x)  \right\} \nonumber\\
&=& 
\sum_{x \in \Gamma} \omega(x) \,{\rm tr}\{ Q \gamma_5(1- D_L) \}(x,x) . 
\end{eqnarray} 

\end{enumerate}

\subsection{Locality property of the measure term } 
The locality property of the current $j_\mu^\diamond(x)$ 
may be examined by following the procedure 
to decompose 
the measure term eq.~(\ref{eq:measure-term-finite-volume}) into the part definable in infinite volume 
and the part of the finite volume corrections:
\begin{equation}
{\mathfrak{L}}_\eta^\diamond = {\mathfrak{K}}_\eta^\diamond  + {\mathfrak{S}}_\eta^\diamond , 
\end{equation}
where 
\begin{eqnarray}
\label{eq:measure-term-finite-volume-infinite-part}
\mathfrak{K}_\eta^\diamond &=& i \int_0^1 ds \, 
{\rm Tr} \left\{ Q_\Gamma \hat P_- [ \partial_s \hat P_-, 
\delta_\eta \hat P_- ] \right\} \nonumber\\
&& \qquad \qquad 
+
\delta_\eta \, 
\int_0^1 ds \, \sum_{x \in \Gamma} 
 \left\{   \tilde A_\mu^\prime(x)  \,   \bar k_\mu(x) \right\} ,  \\
\label{eq:measure-term-finite-volume-finite-part}
\mathfrak{S}_\eta^\diamond &=& \,  
\int_0^1 ds  \, \, 
\mathfrak{R}_{\zeta \eta}\vert_{\zeta_\mu = \tilde A_\mu^\prime} \nonumber\\
&&  \qquad \qquad
+\delta_\eta \, 
\int_0^1 ds \, \sum_{x \in \Gamma} 
 \left\{   \tilde A_\mu^\prime(x)  \,   \Delta k_\mu(x) \right\}   
+{\mathfrak{L}}_\eta \vert_{U=U_{[w]} V_{[m]}, \eta=\eta_{[w]}}. 
\end{eqnarray}
From eqs.~(\ref{eq:bound-R}), (\ref{eq:bound-kmu}), 
(\ref{eq:measure-term-Wilson-lines-bound}) and 
$\| A_\mu^T(x) \| \le \kappa_6 L^4 \, (\kappa_6 >0)$ \cite{Luscher:1998du}, one can infer 
\begin{equation}
\label{eq:bound-S-diamond}
 \left| \mathfrak{S}^\diamond_{\eta} \right|  \le 
\kappa_7 L^{\nu_7} {\rm e}^{-L/\varrho} \, \| \eta \|_\infty 
\end{equation}
for some constants $\kappa_7 >0$, $\nu_7 \ge 0$.

As to ${\mathfrak{K}}_\eta^\diamond$ defined by  
eq.~(\ref{eq:measure-term-finite-volume-infinite-part}), 
if one introduces the truncated fields
\begin{equation}
\eta^n_\mu(x) = \left\{ \begin{array}{cl} \eta_\mu(x) & \text{if $x-Ln \in \Gamma$}, \\
                                                                     0 & \text{otherwise, } \end{array} \right. 
\end{equation}
for any integer vector $n$, 
it may be rewritten into 
\begin{eqnarray}
\label{eq:measure-term-finite-volume-infinite-part-2}
\mathfrak{K}_\eta^\diamond &=& i \int_0^1 ds \, 
{\rm Tr} \left\{ P [ \partial_s P, \delta_{\eta^0} P ] \right\} 
\nonumber\\
&& \qquad \quad 
+
\int_0^1 ds \, \sum_{x \in \mathbb{Z}^4} \left\{ 
( \eta_\mu^0(x)-\eta_{\mu[w]}^0(x) )\, \bar k_\mu(x) 
+  \tilde A_\mu^\prime(x)  \,   \delta_{\eta^0} \, \bar k_\mu(x) \right\}. 
\end{eqnarray}
One can see from this expression
that $\mathfrak{K}_\eta^\diamond$ is defined in infinite volume for  
the variational parameter with a compact support $\eta_\mu^0(x)$. 
Then the following lemma holds ture:

\vspace{1.5em}
\noindent{\bf Lemma 3.d} \ \ {\sl ${\mathfrak{K}}_\eta^\diamond$ is in the form
\begin{equation}
\label{eq:K-to-L-star}
{\mathfrak{K}}_\eta^\diamond = \mathfrak{L}^\star_{\eta^0 [m]}, 
\end{equation}
where $\mathfrak{L}^\star_{\eta [m]}$ is the linear functional defined in infinite volume 
for any variation parameter $\eta_\mu(x)$ with a compact support given by
\begin{eqnarray}
\label{eq:measure-term-infinite-volume-star-m}
\mathfrak{L}_{\eta [m]}^\star &=&\int_0^1 ds \, 
\Big[
 i {\rm Tr} \left\{ P [ \partial_s P, \delta_{\eta} P ] \right\} 
+
\sum_{x \in \mathbb{Z}^4} \Big\{ 
\eta_\mu(x) \, \bar k_\mu(x) 
+  \tilde A_\mu(x)  \,   \delta_{\eta} \, \bar k_\mu(x) \Big\}
\Big]_{U_s={\rm e}^{i  s \tilde A_\mu} \,  V_{[m]}} 
\nonumber\\
&\equiv& \sum_{x \in \mathbb{Z}^4} \eta_\mu(x) j_{\mu [m]}^\star(x) . 
\end{eqnarray}
$\tilde A_\mu(x)$ here 
is the vector potential which represents the dynamical degrees of freedom of the link field in the given topological sector $\mathfrak{U}[m]$,  $\tilde U(x,\mu)=U(x,\mu) V_{[m]}(x,\mu)^{-1}$, 
with the following properties,  
\begin{eqnarray}
\label{eq:prop-tilde-A}
&& U(x,\mu) = {\rm e}^{i \tilde A_\mu(x) } V_{[m]}(x,\mu) ,  
\quad   | \tilde A_\mu(x) |  \le \pi( 1 + 4 \| x \| ) , \nonumber\\
&& F_{\mu\nu}(x) = \partial_\mu \tilde A_\nu(x)- \partial_\nu \tilde A_\mu(x)
+ \frac{2\pi m_{\mu\nu}}{L^2} ,  
\end{eqnarray}
and any other field with these properties is equal to $\tilde A_\mu(x)+\partial_\mu \omega(x)$, 
where the gauge function $\omega(x)$ takes values that are integer multiples of $2\pi$.
}

\vspace{2em}
\noindent
The current $j_{\mu [m]}^\star(x)$ is quite similar in construction 
to $j_\mu^\star(x)$ defined by eq.~(\ref{eq:measure-term-in-infinite-volume})
except the fact that $V_{[m]}(x,\mu)$ is chosen as the reference field at $s=0$. 
In particular, 
$j_{\mu [m]}^\star(x)$ is invariant under the gauge transformations
$\tilde A_\mu(x) \rightarrow \tilde A_\mu(x)+\partial_\mu \omega(x)$ for arbitrary  gauge functions 
$\omega(x)$ that are  polynomially bounded at infinity.  Then, 
the locality property of $j_{\mu [m]}^\star(x)$ can be established by the same 
argument as that given in  \cite{Luscher:1998du}, or in section~\ref{sec:first-step-locality}.

\subsubsection{Proof of the lemma 3.d}
The proof of the lemma 3.d may be given as follows.
By noting eq.~(\ref{eq:wilon-line-in-pure-gauge}), we consider to change the one-parameter 
family of the link fields for the $s$-parameter integration
by the shift of the vector potential, 
\begin{equation}
\tilde A_\mu^\prime(x) \longrightarrow \tilde A_\mu(x) 
= \tilde A_\mu^\prime(x) - \partial_\mu  \Omega_{[w]}(x) \, \, ; \qquad
\Omega_{[w]}(x) = \frac{1}{i}   \ln \Lambda_{[w]}(x), 
\end{equation}
so that the degrees of freedom of the Wilson lines are included in the vector potential.   
Since one may express $U_s(x,\mu)$ as 
\begin{eqnarray}
U_s(x,\mu)
&=& {\rm e}^{i s \tilde A_\mu^\prime(x)} \, 
{\rm e}^{i \Omega_{[w]}(x)} {\rm e}^{-i \Omega_{[w]}(x+\hat\mu)} \, 
V_{[m]}(x,\mu) 
\nonumber\\
&=& {\rm e}^{i s  \tilde A_\mu(x) }  \, 
{\rm e}^{i (1-s) \Omega_{[w]}(x)} {\rm e}^{-i(1-s) \Omega_{[w]}(x+\hat\mu)} \, 
                    V_{[m]}(x,\mu) . 
\end{eqnarray}
Accordingly, we have
\begin{eqnarray}
\partial_s P \vert_{U_s={\rm e}^{i  s \tilde A_\mu^\prime} \,  U_{[w]} \,  V_{[m]}}
&=& 
\Lambda_{[w]}^{(1-s)} \left\{
\partial_s P 
- i  [ \Omega_{[w]} Q , P]
\right\}_{U_s={\rm e}^{i  s \tilde A_\mu} \,  V_{[m]}} \, 
 \Lambda_{[w]}^{-(1-s)},   
\nonumber\\
&&\nonumber\\
\delta_{\eta} P \vert_{U_s={\rm e}^{i  s \tilde A_\mu^\prime} \,  U_{[w]} \, V_{[m]}}
&=&
\Lambda_{[w]}^{(1-s)} \left\{
 \delta_{ \eta} P 
+ i (1-s) [  \delta_{\eta} \Omega_{[w]} Q , P]
\right\}_{U_s={\rm e}^{i  s \tilde A_\mu} \,  V_{[m]}} \, 
 \Lambda_{[w]}^{-(1-s)}.   
\nonumber\\
\end{eqnarray}
Then, the first term in the r.h.s. of eq.~(\ref{eq:measure-term-finite-volume-infinite-part-2}) reads\begin{eqnarray}
\label{eq:infinite-part-evaluation1}
 i \int_0^1 ds \, {\rm Tr} \left\{ P [ \partial_s P, \delta_{ \eta^0} P ] \right\} 
&=&
i \int_0^1 ds \, {\rm Tr} \left\{  P [ \partial_s P, \delta_{\eta^0} P ] \right\} 
  \vert_{U_s={\rm e}^{i  s \tilde A_\mu}  \,  V_{[m]}}
  \nonumber\\
&+& \int_0^1 ds \,   {\rm Tr} \left\{Q_\Gamma P [  [ \Omega_{[w]} Q , P], \delta_{ \eta} P ] ]
\right\} 
  \vert_{U_s={\rm e}^{i  s \tilde A_\mu} \,  V_{[m]}}
\nonumber\\
&-& \int_0^1 ds \,  (1-s) \,  {\rm Tr} \left\{Q_\Gamma P [ \partial_s P, [  \delta_{\eta} \Omega_{[w]} Q , P] ]
\right\} 
  \vert_{U_s={\rm e}^{i  s \tilde A_\mu} \,  V_{[m]}}
\nonumber\\
&+& i  \int_0^1 ds \,  (1-s) \,  {\rm Tr} \left\{Q_\Gamma P 
[ [ \Omega_{[w]} Q , P], [  \delta_{\eta} \Omega_{[w]} Q , P] ]
\right\} 
  \vert_{U_s={\rm e}^{i  s \tilde A_\mu} \,  V_{[m]}} . 
\nonumber\\
\end{eqnarray}
In the r.h.s. of eq.~(\ref{eq:infinite-part-evaluation1}), 
the second term  may be evaluated as follows:
\begin{eqnarray}
 \int_0^1 ds \,   {\rm Tr} \left\{Q_\Gamma P [  [ \Omega_{[w]} Q , P], \delta_{ \eta} P ] ]
\right\} 
&=&
 \int_0^1 ds \,   {\rm Tr} \left\{P [  [ \Omega_{[w]} Q , P], \delta_{ \eta^0} P ] ]
\right\} 
\nonumber\\
&=&
- \int_0^1 ds \,   {\rm Tr} \left\{\Omega_{[w]} Q  \,  \delta_{\eta^0} P\right\} 
\nonumber\\
&=&
\int_0^1 ds \, 
\sum_{x \in \mathbb{Z}^4} 
\left\{ 
(-\partial_\mu \Omega_{[w]}(x) )  \,   \delta_{\eta^0} \, \bar k_\mu(x) 
\right\} .  
\end{eqnarray}
In this evaluation, we should note that although 
$\Omega_{[m]}(x)= \sum_\mu n_\mu \ln w_\mu$ ($x-nL \in \Gamma$) is not periodic, 
the operator $P [  [ \Omega_{[w]} Q , P], \delta_{ \eta} P ] ]$ is translational invariant, 
\begin{equation}
P [  [ \Omega_{[w]} Q , P], \delta_{ \eta} P ] ](x,y) 
= P [  [ \Omega_{[w]} Q , P], \delta_{ \eta} P ] ](x+n_0 L, y+n_0L) 
\end{equation}
for any constant integer vector $n_0$, 
because the shift in $\Omega_{[w]}(x)$, $\Omega_{[w]}(x+n_0L)-\Omega_{[w]}(x)$,  is independent of $x$ 
and does not contribute to the operator. 
As to the third term in the r.h.s. of eq.~(\ref{eq:infinite-part-evaluation1}), we introduce the truncation of the $s$-differential as
\begin{equation}
(\partial_s)^n U_s(x,\mu) = \left\{ \begin{array}{cl} 
i \tilde A_\mu(x) \, U_s(x,\mu)  \, \, & \text{if $x-Ln \in \Gamma$}, \\
0 & \text{otherwise, } \end{array} \right. 
\end{equation}
for any integer vector $n$. Then, it may be evaluated as follows:
\begin{eqnarray}
&& - \int_0^1 ds \,  (1-s) \,  {\rm Tr} \left\{Q_\Gamma P [ \partial_s P, [  \delta_{\eta} \Omega_{[w]} Q , P] ]
\right\} 
\nonumber\\
&&=
- \int_0^1 ds \,  (1-s) \,  {\rm Tr} \left\{ P [ (\partial_s)^0 P, [  \delta_{\eta} \Omega_{[w]} Q , P] ]
\right\} 
\nonumber\\
&&=
- \int_0^1 ds \,   (1-s) \,  {\rm Tr} \left\{\delta_{\eta}\Omega_{[w]} Q \, (\partial_s)^0 P 
\right\} 
\nonumber\\
&&=
\int_0^1 ds \,   (1-s) \,  
\sum_{x\in \mathbb{Z}^4} \,\left\{ 
 \delta_{\eta} (-\partial_\mu \Omega_{[w]}(x) )  \, (\partial_s)^0 \bar k_\mu(x) \right\}
\nonumber\\
&&=
\int_0^1 ds \,   (1-s) \,  
\sum_{x \in \Gamma}  
\eta_{\mu[w]}(x)  \, \partial_s \bar k_\mu(x) 
\nonumber\\
&&=
\int_0^1 ds \, 
\sum_{x \in \mathbb{Z}^4}  
\eta_{\mu[w]}^0(x)  \, \bar k_\mu(x) 
\, \vert_{U_s={\rm e}^{i  s \tilde A_\mu} \,  V_{[m]}} 
-  \sum_{x \in \mathbb{Z}^4}  
\eta_{\mu[w]}^0(x)  \, \bar k_\mu(x) 
\, \vert_{U= V_{[m]}}. 
\end{eqnarray}
In the last two steps above, we have used the relation 
\begin{equation}
\delta_{\eta}  ( - \partial_\mu \Omega_{[w]}(x) )
=(1/i) \, \delta_{\eta} U_{[w]}(x,\mu) \,  U_{[w]}(x,\mu)^{-1} = \eta_{\mu [w]}(x) , 
\end{equation}
and the fact that the local, gauge-invariant current $\bar k_\mu(x)$ at the link field 
$U(x,\mu)=V_{[m]}(x,\mu)$ with the constant field tensor is independent of $x$ and may be set to zero:
\begin{equation}
\left.  \bar k_\mu(x) \right\vert_{U=  V_{[m]}}  \equiv 0 . 
\end{equation}
%
%
The fourth term in the r.h.s. of eq.~(\ref{eq:infinite-part-evaluation1}) turns out to vanish identically:
by noting the hermiticity of $P(x,y)$, this term reads 
\begin{eqnarray}
&&
 i  \int_0^1 ds \,  (1-s) \,  {\rm Tr} \left\{Q_\Gamma P 
[ [ \Omega_{[w]} Q , P], [  \delta_{\eta} \Omega_{[w]} Q , P] ]
\right\} 
\nonumber\\
&&=
 i  \int_0^1 ds \,  (1-s) \, \frac{1}{2} {\rm Tr} \left\{Q_\Gamma \left(
   P \Omega_{[w]} Q P \delta_{\eta} \Omega_{[w]} Q P 
 - P \delta_{\eta}\Omega_{[w]} Q P  \Omega_{[w]} Q P     \right. \right.
 \nonumber\\
 && \qquad \qquad\qquad \qquad\qquad\qquad
 \left. \left.
 - P \Omega_{[w]} Q \, \delta_{\eta}\Omega_{[w]} Q P 
 +P \delta_{\eta}\Omega_{[w]} Q \,  \Omega_{[w]} Q P \right)
\right\}
\nonumber\\
&&= 
 i  \int_0^1 ds \,  (1-s) \, \frac{1}{2} \,  \sum_{\mu \nu} [(1/i) \ln w_\mu] \, \eta_{(\nu)} \, 
\left[ I_{\mu\nu} - I_{\nu\mu} - J_{\mu\nu} + J_{\nu\mu} \right], 
\end{eqnarray}
where 
\begin{eqnarray}
I_{\mu\nu} &=&  \sum_{x,y,z \in\Gamma} \sum_{n,n^\prime \in \mathbb{Z}^4} 
{\rm tr} \left\{ P(x, y+nL) n_\mu P(y+nL,z+n^\prime L ) n_\nu^\prime P(z+n^\prime L,x) \right\}  , \\
J_{\mu\nu} &=&  \sum_{x,y \in\Gamma} \sum_{n \in \mathbb{Z}^4} 
{\rm tr} \left\{ P(x, y+nL) n_\mu n_\nu P(y+n L ,x) \right\} ,  
\end{eqnarray}
but, $I_{\mu\nu}$ and $J_{\mu\nu}$ are both symmetric with respect to the indices $\mu, \nu$. 
Combining these results and 
using the gauge invariance of $\bar k_\mu(x)$, 
\begin{eqnarray}
 &&
\bar k_\mu(x) \vert_{U_s={\rm e}^{i  s \tilde A_\mu^\prime} \,  V_{[m]}} 
= \bar k_\mu(x)  \vert_{U_s={\rm e}^{i  s \tilde A_\mu} \,  V_{[m]}} , 
\end{eqnarray}
in  the second term of the r.h.s. of eq.~(\ref{eq:measure-term-finite-volume-infinite-part-2}),  
we finally obtain 
\begin{eqnarray}
\label{eq:measure-term-finite-volume-infinite-part-3}
\mathfrak{K}_\eta^\diamond &=&\int_0^1 ds \, 
\Big[
 i {\rm Tr} \left\{ P [ \partial_s P, \delta_{\eta^0} P ] \right\} 
+
\sum_{x \in \mathbb{Z}^4} \left\{ 
\eta_\mu^0(x) \, \bar k_\mu(x) 
+  \tilde A_\mu(x)  \,   \delta_{\eta^0} \, \bar k_\mu(x) \right\}
\Big]_{U_s={\rm e}^{i  s \tilde A_\mu} \,  V_{[m]}} . 
\nonumber\\
\end{eqnarray}

As the final step, we cast the vector potential $\tilde A_\mu(x)$, 
which represents 
\begin{equation}
\tilde U(x,\mu)
={\rm e}^{i A_\mu^T(x)} \Lambda(x) \, \Lambda(x+\hat \mu)^{-1} \, U_{[w]}(x,\mu) 
= U(x,\mu) V_{[m]}(x,\mu)^{-1} , 
\end{equation}
the dynamical degrees of freedom of the link field in the given topological sector $\mathfrak{U}[m]$,  
into the complete axial gauge 
with the properties eq.~(\ref{eq:prop-tilde-A}), 
by applying the same construction as in the lemma 2.b 
\cite{Luscher:1998kn}  to $\tilde U(x,\mu)$. 
Since $\mathfrak{K}_\eta^\diamond$ is invariant under the gauge transformation 
$\tilde A_\mu(x) \rightarrow \tilde A_\mu(x)+\partial_\mu \omega(x)$
for arbitrary  gauge functions 
$\omega(x)$ that are  polynomially bounded at infinity, 
as easily verified, 
this change of the gauge for $\tilde A_\mu(x)$ does not alter $\mathfrak{K}_\eta^\diamond$ itself. 
This results in eq.~(\ref{eq:K-to-L-star}) of the lemma 3.d. 






\section{Discussion}
\label{sec:discussion}

We have given a closed formula eq.~(\ref{eq:measure-term-finite-volume}) 
of the measure term on the finite volume lattice, which fulfills 
all the required properties for the reconstruction theorem
in the gauge-invariant formulation of U(1) chiral gauge theories \cite{Luscher:1998du}. 
Although it is intended for the use in a practical implementation of U(1) chiral lattice gauge theories, 
it also provides, we believe,  a simpler point of view on the theoretical structure of the formulation. 

A comment is in order about the relation between 
the measure term $\mathfrak{L}^\diamond_{\eta}$ constructed in this paper and 
the measure  term  $\mathfrak{L}_{\eta}$ given in the original 
construction \cite{Luscher:1998du}.  
Since both terms satisfy the integrability condition and the anomalous conservation law, 
one may expect that they are related each other by 
the variation of a certain gauge-invariant local term as 
\begin{equation}
\label{eq:relation-Ldiamond-Loriginal}
\mathfrak{L}^\diamond_{\eta}= \mathfrak{L}_{\eta} 
+ \sum_{x\in \Gamma} \delta_\eta \mathfrak{D}(x) .  
\end{equation}
In fact, as to the terms definable in infinite volume,  
it is possible to work out the difference between the linear functionals 
$\mathfrak{K}^\diamond_{\eta^0}=\mathfrak{L}^\star_{\eta^0 [m]}$ and 
$\mathfrak{K}_{\eta^0}=\mathfrak{L}^\star_{\eta^0}$
explicitly and the result is given in the following form:
\begin{equation}
\mathfrak{L}^\star_{\eta^0 [m]}= \mathfrak{L}^\star_{\eta^0} 
+ \sum_{x\in \mathbb{Z}^4} \delta_{\eta^0} \mathfrak{D}^\star_{[m]}(x) , 
\end{equation}
where $\mathfrak{D}^\star_{[m]}(x)$ is the local field given by 
\begin{eqnarray}
\mathfrak{D}^\star_{[m]}(x)
&=&\int_0^1 dt \, \int_0^1 ds \, \left[  
i\,   {\rm tr}\left\{ P[\partial_t P, \partial_s P ] \right\}(x,x)  \phantom{\tilde A_\mu(x)}
\right.
\nonumber\\
&& 
\left. + (s \tilde A_\mu + [A_\mu(x)-\tilde A_\mu(x)] ) \partial_s \bar k_\mu(x) 
         - t \tilde A_\mu(x) \partial_t \bar k_\mu(x)  
\right]_{U_{t,s}={\rm e}^{i t ( s \tilde A + [A-\tilde A] )} } . 
\end{eqnarray}
The relation eq.~(\ref{eq:relation-Ldiamond-Loriginal}) implies that the resulted 
Weyl fermion measures,  or the effective actions induced by the Weyl fermion path integral, 
differ by the  gauge-invariant local term $\sum_{x\in \Gamma}  \mathfrak{D}(x)$. 
We do not know, however,  if there exists a closed expression of $\mathfrak{D}(x)$ in terms of only 
the quantities defined in finite volume like 
eq.~(\ref{eq:measure-term-finite-volume}) for $\mathfrak{L}^\diamond_{\eta}$.

In the formula of the measure term, eq.~(\ref{eq:measure-term-finite-volume}), we have adopted the transverse gauge for the vector potential 
$\tilde A^\prime_\mu(x)=A_\mu^T(x) - \frac{1}{i}\partial_\mu \Big[ \ln \Lambda(x) \Big] $ such that
\begin{eqnarray}
{\rm e}^{i \tilde A^\prime_\mu(x) } 
&=&  {\rm e}^{i  A^T_\mu(x) } \Lambda(x) \Lambda(x+\hat\mu)^{-1}  \nonumber\\
&=& U(x,\mu)  \, V_{[m]}(x,\mu)^{-1} \, U_{[w]}(x,\mu)^{-1}
\equiv \tilde U^\prime(x,\mu) . 
\end{eqnarray}
Since the measure term (current) is gauge-invariant, one may choose different gauge conditions. 
For example, one may adopt the complete axial gauge, inspired by the following lemma ($L$ is assumed to be an even number):

\vspace{1em}
{\sl
\noindent{\bf Lemma 4.a} \ \ There
exists a periodic vector potential
$\tilde A_\mu^\prime(x)$  such that 
\begin{eqnarray}
 {\rm e}^{i \tilde A_\mu^\prime(x)}&=& \tilde U^\prime(x,\mu) , \\
\partial_\mu \tilde A_\nu^\prime(x) - \partial_\nu \tilde A_\mu^\prime(x) 
&=&  F_{\mu\nu}(x)- \frac{2\pi m_{\mu\nu}}{L^2}, \\
 \sum_{s=1}^L \tilde A_\mu^\prime(x+s \hat \mu) &=& 
\frac{1}{i} \, \ln \left[ \prod_{s=1}^L \tilde U^\prime(x+s \hat \mu,\mu)  \right]  , 
\end{eqnarray}
and
\begin{equation}
\left\{ 
\begin{array}{ll}
\vert \tilde A_\mu^\prime(x) \vert \le \pi(1+4 \| x -x_0 \|) &\ \ 
{\rm for} \ \ \vert (x-x_0)_\nu \vert \not = L/2-1  \ (\nu=1,2,3,4),\\
\vert \tilde A_\mu^\prime(x) \vert \le \pi(1 + 6L^2 +2L( 1+ 3 L^2) ) & \ \ {\rm otherwise} . 
\end{array}
\right.
\end{equation}
$x_0$ is a reference point which may be chosen arbitrarily. 
Moreover, if $\tilde A_\mu^{\prime\prime}(x)$ is any other
field with these properties we have
\begin{equation}
\label{eq:gauge-equivalence}
\tilde A_\mu^{\prime\prime}(x) = \tilde A_\mu^\prime(x)+ \partial_\mu 
\omega(x), 
\end{equation}
where the gauge function $\omega(x)$ is {\em periodic} and takes values that are 
integer multiples of $2\pi$. 
}

\vspace{1em}
\noindent
The proof of this lemma is given in appendix~\ref{appendix:vector-potential}.
Note that
the variation of the vector potential in this gauge is also given by 
$\delta_\eta \tilde A^\prime_\mu(x) = \eta_\mu(x)-\eta_{\mu [w]}(x)$. 

Given the  local current $j_\mu^\diamond(x)$,  the basis vectors of the Weyl field 
can be constructed explicitly as follows \cite{Luscher:1999un}:
\begin{equation}
\label{eq:chiral-basis-choice}
v_j(x) = \left\{    \begin{array}{ll}
   Q_1  v_{1 [m]} \, W^{-1} &\quad \text{if $j=1$} , \\
   Q_1  v_{j [m]}                &\quad \text{otherwise} , 
   \end{array} \right.
\end{equation}
where,  
along the one-parameter family of the link fields in $\mathfrak{U}[m]$,  
\begin{equation}
  U_t(x,\mu)= {\rm e}^{i t \tilde A_\mu^\prime(x)} \,  U_{[w]}(x,\mu) \, 
  V_{[m]}(x,\mu),\qquad   0 \le t \le 1 , 
\end{equation}
$W$ is defined by
\begin{equation}
W \equiv {\rm exp}\left\{ i \int_0^1 dt  \, \mathfrak{L}^\diamond_\eta \right\}, 
   \qquad
\eta_\mu(x) = i \partial_t U_t(x,\mu) \, U_t(x,\mu)^{-1} , 
\end{equation}
$Q_t$ is defined by the evolution operator of the projector
$P_t = \left. \hat P_- \right\vert_{U=U_t}$ satisfying 
\begin{equation}
\partial_t Q_t = \left[ \partial_t P_t , P_t \right] Q_t ,   \quad Q_0 = 1 , 
\end{equation}
and 
$v_{j [m]}$ are the basis vectors at the Wilson lines 
$U_{[w]}(x,\mu)V_{[m]}(x,\mu)$ ($t=0$).\footnote{
With this definition, the explicit expression of $W$ is given by
\begin{equation}
W = \exp \left\{ 
i \int_0^1 ds \,    \sum_{x\in \Gamma} 
\tilde A_\mu^\prime(x)  \,  \left.  k_\mu(x) \right\vert_{\tilde A^\prime \rightarrow s \tilde A^\prime}   \right\} . 
\end{equation}
}
Towards a numerical  application of U(1) chiral lattice gauge theories, 
a next step is the practical implementation of this formula: a computation of $W$, 
the implementation of the operator $Q_t$ and the construction of $v_{j [m]}$. 
This question has been addressed partly in our previous 
works \cite{Kadoh:2004uu, Kikukawa:2001mw, Aoyama:1999hg}. 
We will disscuss this question in full detail elsewhere.

\bigskip

\acknowledgments

The authors would like to thank M.~L\"uscher for valuable comments.
Y.K. is grateful to Ting-Wai Chiu for his kind hospitality at 2005 Taipei
Summer Institute on Strings, Particles and Fields and 
D.~Adams, K.~Fujikawa,  H.~Suzuki for discussions. 
Y.K. is supported in part by Grant-in-Aid for Scientific Research No.~17540249.

\bigskip
\appendix

\section{Proof of the lemma 4.a }
\label{appendix:vector-potential}

For simplicity, we set the reference point to the origin, $x_0=0$. 
The extention of the following proof to the case 
with a generic reference point $x_0$ is straightforward. 

We introduce a vector potential 
\begin{equation}
\tilde a_\mu(x) = \frac{1}{i} 
\ln \left[   
 \tilde U^\prime(x,\mu)     
\right], \qquad
-\pi < \tilde a_\mu(x) \le \pi
\end{equation}
where
\begin{equation}
\tilde U^\prime(x,\mu) 
= {\rm e}^{i A^T_\mu (x)} \, \Lambda(x) \Lambda(x+\hat\mu)^{-1} 
=  U(x,\mu) \, V_{[m]}(x,\mu)^{-1} \,  U_{[w]}(x,\mu)^{-1} , 
\end{equation}
and then note that
\begin{equation}
F_{\mu\nu}(x)=\partial_\mu \tilde a_\nu(x) - \partial_\nu 
\tilde a_\mu(x) 
+ \frac{2\pi m_{\mu\nu}}{L^2}
+ 2\pi \tilde n_{\mu\nu}(x),
\end{equation}
where $\tilde n_{\mu\nu}(x)$ is an anti-symmetric tensor
field with integer values which satisfies 
\begin{eqnarray}
&& \partial_{[\rho} \tilde n_{\mu\nu]}(x)= 0 , \\
&& \sum_{s,t=0}^{L-1} \tilde n_{\mu\nu}(x+s\hat\mu+t\hat\nu)=0.
\end{eqnarray}
The Bianci identity of $\tilde n_{\mu\nu}(x)$ follows from 
the Bianci identity of $F_{\mu\nu}(x)$ which holds true for 
$\epsilon < \pi/3$.

We now construct a {\it periodic} integer vector field 
$\tilde m_\mu(x)$ such that 
$\partial_\mu \tilde m_\nu-\partial_\nu \tilde m_\mu = \tilde
n_{\mu\nu}$. For this purpose, we try to impose a complete axial 
gauge where
$\tilde m_1(x)=0$, 
$\tilde m_2(x)\vert_{x_1=0}=0$, 
$\tilde m_3(x)\vert_{x_1=x_2=0}=0$, 
$\tilde m_4(x)\vert_{x_1=x_2=x_3=0}=0$ 
and to obtain the non-zero components of the field by solving
\begin{equation}
\partial_\mu \tilde m_\nu(x) = \tilde n_{\mu\nu}(x) \ \  at \ \ 
x_1=\cdots = x_{\mu-1} = 0 
\end{equation}
for $\mu=3,2,1$ (in this order) and $\nu > \mu$. However, 
the resulted  
vector potential is not periodic.
Let us denote the restriction of the solution on to $\Gamma$
by $m_\mu(x)$,
\begin{equation}
m_\mu(x)
= 
- \sum_{\nu < \mu} \sum_{t_\nu=0}^{x_\nu-1}{}^\prime \, 
\left.  \tilde n_{\mu\nu}(z^{(\nu)}) \right \vert_{x_1=\cdots=x_{\nu-1}=0}
\end{equation}
where $x \in \Gamma$, 
$z^{(\nu)}=(x_1,\cdots,t_\nu,\cdots)$
and 
\begin{equation}
\sum_{t_i =0}^{x_i-1}{}^\prime f(x)
=\left\{ 
\begin{array}{ll}
\sum_{t_i=0}^{x_i-1} f(x) & (x_i \ge 1 )\\
0 & (x_i =0 )\\
\sum_{t_i=x_i}^{-1} (-1) f(x) & (x_i \le -1 )
\end{array}
 \right. .
\end{equation}
Although it satisfies the bound $\vert m_\mu(x) \vert \le
2 \parallel x \parallel$, it only satisfies
\begin{equation}
\tilde n_{\mu\nu}= \partial_\mu m_\nu-\partial_\nu m_\mu 
+ \Delta \tilde n_{\mu\nu},
\end{equation}
\begin{equation}
\Delta \tilde n_{\mu\nu}(x)= 
\delta_{x_\mu,L/2-1}
\sum_{\tilde t_\mu=0}^{L-1}
\left.  \tilde n_{\mu\nu}(z^{(\mu)}) \right \vert_{x_1=\cdots=x_{\mu-1}=0},
%
%
\end{equation}
where $\nu > \mu$ and 
$\tilde t_\mu = t_\mu$ mod $L$. 
We note that $\Delta \tilde n_{\mu\nu}(x)$ has the support on the 
boundary of $\Gamma$. We then use the lattice counterpart
of the lemma 9.2 in \cite{Luscher:1998du}, to obtain the
periodic  integer vector potential $\Delta m_\mu(x)$ which solve
$\partial_\mu \Delta m_\nu-\partial_\nu \Delta m_\mu = 
\Delta \tilde n_{\mu\nu}$, 
\begin{eqnarray}
\Delta m_\mu(x) =\hspace{6cm}\nonumber\\
-
\delta_{x_\mu,L/2-1}
\sum_{\nu > \mu}
\sum_{t_\nu=0}^{x_\nu-1}{}^\prime 
\sum_{\tilde t_\mu=0}^{L-1} \, 
\left.  \tilde n_{\mu\nu}(z^{(\mu,\nu)}) \right \vert_{x_{\nu+1}=\cdots=0},
\end{eqnarray}
The desired periodic integer vector potential 
$\tilde m_\mu(x)$ is now obtained by 
$\tilde m_\mu(x) = m_\mu(x) + \Delta m_\mu(x)$, 
which satisfies the bound
\begin{equation}
\left\{ 
\begin{array}{ll}
\vert \tilde m_\mu(x) \vert \le 2 \parallel x \parallel &
{\rm for} \ \ x_\nu \not =  L/2-1  \ (\nu=1,2,3),\\
\vert \tilde m_\mu(x) \vert \le 3L^2 & {\rm otherwise}.
\end{array}
\right.
\end{equation}

Finally, we note that the differences of the Wilson lines between $\tilde U^\prime(x,\mu)$ and 
$\tilde a_\mu(x) + 2 \pi \tilde m_\mu(x)$ are integer multiples of $2\pi$. Namely, one has
\begin{eqnarray}
 \frac{1}{i} \, \ln \left[ \prod_{s=1}^L \tilde U^\prime(x+s \hat \mu,\mu)  \right]
- \sum_{s=1}^L  \left\{ \tilde a_\mu(x+s \hat \mu) + 2 \pi \tilde m_\mu(x+s \hat \mu) \right\}  
= 2 \pi c_\mu(x), 
\end{eqnarray}
where $c_\mu(x) (\mu=1,2,3,4)$ take integer values. One can also infer  the bound
\begin{equation}
| c_\mu(x) |  \, \le \,  L ( 1+ 3 L^2) . 
\end{equation}
Then, the
vector potential with the desired properties  is obtained by
\begin{equation}
\label{eq:vector-potential-finite-lattice}
\tilde A_\mu^{\prime}(x) = \tilde a_\mu(x) + 2 \pi \tilde m_\mu(x) +  \delta_{x_\mu,L/2-1} \,  2 \pi c_\mu(x) . 
\end{equation}

If there exists a vector potential $\tilde A_\mu^{\prime\prime}(x)$ 
with the same properties as $\tilde A_\mu^{\prime}(x)$, 
the vector potential defined by the difference 
$\tilde A_\mu^{\prime\prime}(x)-\tilde A_\mu^{\prime}(x)$ has
the vanishing field tensor and the vanishing Wilson lines.  Such a vector potential should be in pure gauge form, 
\begin{equation}
\label{eq:gauge-equivalence}
\tilde A_\mu^{\prime\prime}(x) -\tilde A_\mu^\prime(x)= \partial_\mu 
\omega(x), 
\end{equation}
where the gauge function $\omega(x)$ is {\em periodic} and takes values that are 
integer multiples of $2\pi$.


\end{document}